\documentclass[pdflatex,sn-mathphys-num]{sn-jnl}%

\usepackage{geometry}
\usepackage{multirow}%
\usepackage{amsmath,amssymb,amsfonts}%
\usepackage{amsthm}%
\usepackage{mathrsfs}%
\usepackage[title]{appendix}%
\usepackage{xcolor}%
\usepackage{textcomp}%
\usepackage{manyfoot}%
\usepackage{booktabs}%
\usepackage{algorithm}%
\usepackage{microtype}
\usepackage{algorithmicx}%
\usepackage{algpseudocode}%
\usepackage{listings}%
\usepackage{amsmath}
\usepackage{amsfonts}
\usepackage{comment}
\usepackage{forest}
\usepackage{amssymb}
\usepackage{graphicx}
\usepackage{tikz}
\usetikzlibrary{shapes.misc}
\usepackage{booktabs} %
\usepackage{adjustbox}
\usepackage{float}
\usepackage{lineno}

\theoremstyle{thmstyletwo}%

\theoremstyle{thmstylethree}%
\usepackage{color}

\usepackage[utf8]{inputenc}
\usepackage{newunicodechar}
\newunicodechar{׳}{'}
\begin{document}

\title[Article Title]{Artificial intelligence is creating a new global linguistic hierarchy}

\author[1]{\fnm{Giulia} \sur{Occhini}}

\author[1,2]{\fnm{Kumiko} \sur{Tanaka-Ishii}}

\author[1]{\fnm{Anna} \sur{Barford}}

\author[1,3]{\fnm{Refael}
\sur{Tikochinski}}

\author[1]{\fnm{Songbo} \sur{Hu}}

\author[1,4]{\fnm{Roi} \sur{Reichart}}

\author[1]{\fnm{Yijie} \sur{Zhou}}

\author[1]{\fnm{Lester James V.} \sur{Miranda}}

\author[1]{\fnm{Hannah} \sur{Claus}}

\author[1]{\fnm{Ulla} \sur{Petti}}

\author[1]{\fnm{Ivan} \sur{Vulic}}

\author[1]{\fnm{Ramit} \sur{Debnath}}

\author[1]{\fnm{Anna} \sur{Korhonen}}

\affil[1]{\orgname{University of Cambridge}, \country{UK}}

\affil[2]{\orgname{Waseda University},  \country{Japan}}

\affil[3]{\orgname{University College London},  \country{UK}}

\affil[4]{\orgname{Technion}, \country{Israel}}

\abstract{Artificial intelligence (AI) has the potential to transform healthcare, education, governance and socioeconomic equity, but its benefits remain concentrated in a small number of languages \citep{bender2019benderrule, joshi2020state, blasi-etal-2022-systematic, young2015digital, ranathunga-de-silva-2022-languages}. 
Language AI, the technologies that underpin widely-used conversational systems such as ChatGPT, could provide major benefits if available in people’s native languages, yet most of the world’s 7,000+ linguistic communities currently lack access and face persistent digital marginalization. Here we present a global longitudinal analysis of social, economic and infrastructural conditions across languages to assess systemic inequalities in language AI. We first analyze the existence of AI resources for 6003 languages. We find that despite efforts of the community to broaden the reach of language technologies \citep{bapna2022building, costa2022no}, the dominance of a handful of languages is exacerbating disparities on an unprecedented scale, with divides widening exponentially rather than narrowing. Further, we contrast the longitudinal diffusion of AI with that of earlier IT technologies, revealing a distinctive hype-driven pattern of spread. To translate our findings into practical insights and guide prioritization efforts, we introduce the Language AI Readiness Index (EQUATE), which maps the state of technological, socio-economic, and infrastructural prerequisites for language technology development across languages. The index highlights communities where capacity exists but remains underutilized, and provides a framework for accelerating more equitable diffusion of language technology. Our work contributes to setting the baseline for a transition towards more sustainable and equitable language technologies.}

\keywords{Digital Language Inequalities, Artificial Intelligence, Technological Development; Social justice}

\maketitle

\section{Introduction}
The rapid advancement of Artificial Intelligence (AI) has brought language technologies, especially conversational systems, to the forefront of innovation. These technologies are widely expected to transform communication, information access, and service delivery across sectors such as healthcare, education, government, and scientific research. However, despite their transformative potential, their benefits remain highly concentrated among a small fraction of the world’s languages and their speakers. As a result, the vast majority of the world’s linguistic diversity, amounting to more than 90\% of living languages, continues to face digital marginalization, with limited or no access to AI-driven tools and resources \citep{joshi2020state, blasi-etal-2022-systematic, ranathunga-de-silva-2022-languages}.

This persistent digital language divide in AI has the potential to determine whether billions of people can access services, information, and education in a language they understand and in a culturally appropriate manner \citep{bromham2022global}. Excluding most languages from digital ecosystems risks to exacerbate inequalities, erode linguistic heritage, and undermine global efforts to promote inclusive and sustainable development \citep{angelo2019well}. International initiatives, including the United Nations (UN)’ Decade of Indigenous Languages (2022–2032), underscore the urgency of safeguarding linguistic diversity and ensuring equitable access to digital technologies \citep{UNESCO_OER_2019}. UNESCO also highlights that 40\% of the global population lacks access to education in a language they speak or understand \citep{unLinguisticDiversity}. For these communities, AI's potential to advance the UN Sustainable Development Goals (SDGs), particularly in reducing inequality and promoting inclusive knowledge access, remains largely unfulfilled \citep{tomavsev2020ai, ftLetterGlobal}. This is in contrast with UN's declaration on rights of indigenous people, having a recognised right to ``revitalise, use develop and transmit" their own language, as well as obtaining education in that language \citep{undrip2007}.

Efforts to broaden AI’s linguistic coverage are growing, as reflected in the proliferation of datasets and models on platforms such as Hugging Face\footnote{https://huggingface.co/}. Yet these initiatives are largely guided by industrial and institutional top-down decisions about which languages to prioritize and developers personal preferences \footnote{Regarding developers' personal preferences, we note how Klingon, an artificial language created for a TV series with 0 speakers, received its first Microsoft translation system already in May 2013 (see, \url{https://www.microsoft.com/en-us/translator/blog/2013/05/14/announcing-klingon-for-bing-translator/}). At that point in time, official national languages with tens of millions of speakers such as Somali, Cantonese and Igbo had no machine translation systems yet.}, often neglecting differences in the technological readiness of specific communities. Even efforts explicitly aimed at democratizing AI by extending its linguistic reach (e.g. \citet{bapna2022building, costa2022no}) remain constrained by the dominance in performance of high-resource languages, so that under-resourced languages are served poorly in the critical domains where it matters most, such as healthcare and education \citep{kobis2025delegation, Kreutzer, wu2020all, bird-2020-decolonising}. This “representation washing” \citep{Kreutzer} inflates linguistic coverage statistics, and risks deepening inequalities rather than reducing them. 

Here, we examine the systemic factors underlying global disparities in language AI. Earlier studies conducted before the emergence of large language models (LLMs) focused mainly on language resources, research coverage and number of speakers \citep{joshi2020state, blasi-etal-2022-systematic}. Inequality in language coverage, however, is influenced by a variety of socio-technical factors foundational to technology itself \citep{nigatu2024zeno}. We take, for the first time, a system-level view of the contemporary language AI era at scale, linking a diverse set of language resources to their socioeconomic and digital-infrastructure contexts across 6,003 languages with at least one documented vocabulary item.

Our findings show a sharp concentration of language technology in only a small set of languages, driven by cumulative advantages that exceed even those predicted by Zipfian distributions. A longitudinal analysis of archival internet data, covering the entire period from the emergence of large scale generative systems to the present day, shows that these disparities have not narrowed; instead, they have intensified over time.
Further, we reveal that, in contrast to classical models of technological diffusion, language technologies spread across languages through hype cycles marked by rapid, uneven surges rather than gradual, steady adoption.

To address these challenges, we introduce the \textit{Language AI Readiness Index (EQUATE)}, a comprehensive metric designed to quantify the readiness of all attested languages and their associated speaker communities for language AI development, evaluated across multiple geographic scales. In contrast to prior readiness assessments, which primarily quantify general-purpose AI capacity at the national level, our approach targets language technologies specifically and operationalizes readiness at the community level, where the most consequential variation for linguistic ecosystems is observed \citep{maslej2025artificial, TortoiseMedia_GlobalAIIndex_2024, oxfordinsights2024Government}.We demonstrate that the selected indicators exhibit joint predictive validity for the trajectory of language AI development, with socioeconomic and infrastructural conditions exerting an independent contribution beyond the influence of pre-existing AI resources. Furthermore, we calibrate the relative weighting of each dimension on the basis of responses obtained from a global survey of domain experts.

By shifting the focus from states to languages and their speaker communities, our EQUATE index provides systematic evidence to guide the development of more equitable language technologies, integrating AI resources, digital infrastructure, and socioeconomic conditions. In countries such as India, for example, where approximately 456 languages are actively spoken \citep{ethnologueWhatCountries}, national and local governments can use the index to guide resource allocation. 

By integrating 29 language-level features, the index provides researchers, developers, policymakers, and other stakeholders with a shared evidence base for identifying priorities and directing resources where they are most needed. It supports more transparent, inclusive, and equitable decision-making about which AI tools and infrastructures to develop, and for whom.

\section{Zipfianisation: Inequalities in language technologies are widening exponentially}

We begin by examining empirical evidence on the ways in which AI-driven language technologies are shaping global linguistic ecologies. Rather than restricting our focus to large, general-purpose language models, we consider the full range of approaches to language AI, including smaller and task-specific models, as well as other established computational approaches to speech and language processing.

To this end, we analyze the distribution of resources on Hugging Face, the predominant platform for hosting models and datasets since its inception in 2020. To capture temporal dynamics, we construct a longitudinal dataset from monthly snapshots obtained via the Wayback Machine\footnote{https://web.archive.org/web/20250000000000*/https://huggingface.co/} for December of each year between 2020 and 2025. We justify this data selection through multiple lines of evidence: specifically, we assess the representativeness of the Hugging Face collection by comparing the number of resources per language with the number of papers per language published in the ACL Anthology, a corpus encompassing all computational linguistics and Natural Language Processing (NLP) papers over the past five decades \citep{radev2013acl} (see Methods in Section \ref{acl_method} and Supplementary Figure \ref{fig:valiud}). Our analysis demonstrates that both the number of language models per language and the volume of associated data follow a power-law distribution, with resource availability declining steeply as language rank increases (see Methods in Section \ref{stat_mod} and Supplementary Table \ref{tab:residuals_std_dev_combined}). This pattern, characterized by an exponent $\alpha$ \citep{newman2005power}, indicates a highly concentrated ecosystem in which a small subset of global lingua francas, primarily English, Mandarin, French, and Spanish, disproportionately dominates technological development.

\begin{figure}
    \centering
    \includegraphics[width=1.0\linewidth]{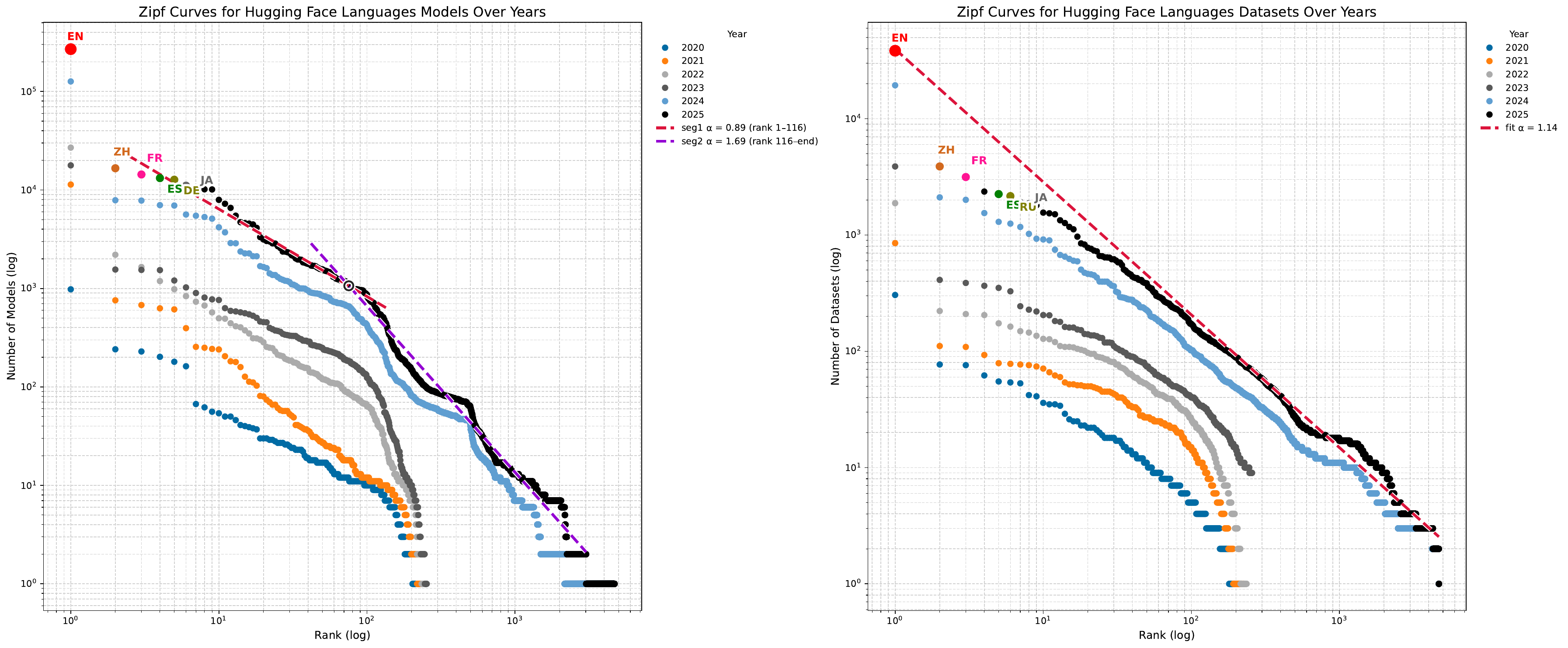}
    \caption{\textbf{Evolution of Language Model and Dataset Distribution on Hugging Face over time (2020-2024) based on Wayback Machine data (Source: \citep{HuggingFace})}. These log-log plots illustrate power-law fits to the 2025 snapshot, showing the frequency of models (left) and datasets (right) per language rank over time. Models follow a broken power law: a shallower head ($\alpha=0.89$) steepens to a heavier tail ($\alpha=1.69$) at a crossover near rank 116.  We observe how the distribution of resources per language reached a Zipf distribution in record time. English (red dot) is a major outlier, with an order of magnitude more models than expected, indicating extreme dominance. Other high-resource languages like French, Spanish, and Chinese follow the Zipf trend more closely. 
}
    \label{fig:longitudinal_zipf}
\end{figure}

The model distribution follows a broken power law rather than a single exponent, with a lower-exponent regime ($\alpha = 0.89$) for the top-ranked languages which steepens ($\alpha = 1.69$) beyond a crossover at rank 116, corresponding to Maithili (see Fig \ref{fig:longitudinal_zipf}).The languages above this point are, with few exceptions, official languages of states or international bodies, backed by established administrative, educational, and publishing infrastructure. Consequently, this crossover point represents more than a statistical discontinuity: it marks a sociolinguistic threshold distinguishing languages with institutional support from those lacking it.

A model's presence on Hugging Face records that development effort was directed at a language, not that the resulting system is functional or high-quality. Our metric therefore captures where effort has been invested and where the conditions for it exist, not the quality of individual models, which no measure at this scale can establish. We subject our findings to two distinct sensitivity analyses. First, we cross-reference model availability against benchmark coverage recovered from the ACL Anthology and find that the observed concentration persists when constrained to an evaluation-based measure derived independently of Hugging Face. We describe how benchmark data were collected and report the full results in Appendix \ref{data} and Supplementary Figure \ref{fig:bench}, respectively. Second, because most contemporary systems are multilingual and a single model may nominally declare dozens or hundreds of languages \citep{Kreutzer}, apparent coverage in the long tail could be sustained by a handful of massively multilingual systems rather than by dedicated resources. We test for this by recomputing resource distribution over monolingual models and datasets alone and find the concentration undiminished. Unsurprisingly, we observe that removing multilingual systems further increases the concentration, with the Gini coefficient rising from 0.910 to 0.966. We report the full results and describe how the monolingual subset was identified in \ref{data} and Supplementary Figure \ref{fig:monol}.

Longitudinal analysis of web archival data over the past five years demonstrates that these disparities are intensifying at an unprecedented rate. We name this process \textit{Zipfianisation}, as illustrated in Figure \ref{fig:longitudinal_zipf}. While English has long been an outlier, its dominance has surged during the LLM era, with model growth rates exceeding even Zipfian predictions (for a comparison of the position of English before the advent of large language models and today, see the Supplementary Figure). For example, English saw an annual increase of up to 50,000 new models, dwarfing gains in other high-resource languages such as Mandarin Chinese. This hyper-concentration suggests a self-reinforcing cycle: languages with existing digital infrastructure and postcolonial status attract disproportionate investment, further entrenching their advantage.

The stability of resource distribution over time underscores the systemic nature of these inequalities. High-resource languages consistently occupy the top ranks, while underrepresented languages, such as Congo Swahili (DRC), Kuanyama (Namibia/Angola), Wolaytta (Ethiopia), and Kwangali (Namibia) remain trapped at the bottom. On average, these languages gain only 4.15 new models annually, compared to the exponential growth in dominant languages. English, in particular, has reached a state of \textit{overresourcing}: it is already supported by a large number of high-capacity systems, so additional English models add little to its coverage. By contrast, equivalent effort directed toward currently unserved languages would generate new capabilities where none presently exist.

This prevailing “winner-takes-all” dynamic not only systematically marginalises the majority of the world’s linguistic diversity, but also risks reproducing and entrenching colonial power structures within digital environments. The implications are unambiguous: in the absence of targeted, sustained intervention, the disparity between over-resourced and under-resourced languages is likely to expand further, thereby exacerbating global inequalities in access to information, public and private services, and educational opportunities.

\newpage

\subsection{Inequality in language technology is geographically shaped}

\begin{figure}[h!]
    \centering
    \includegraphics[width = 1.0\textwidth]{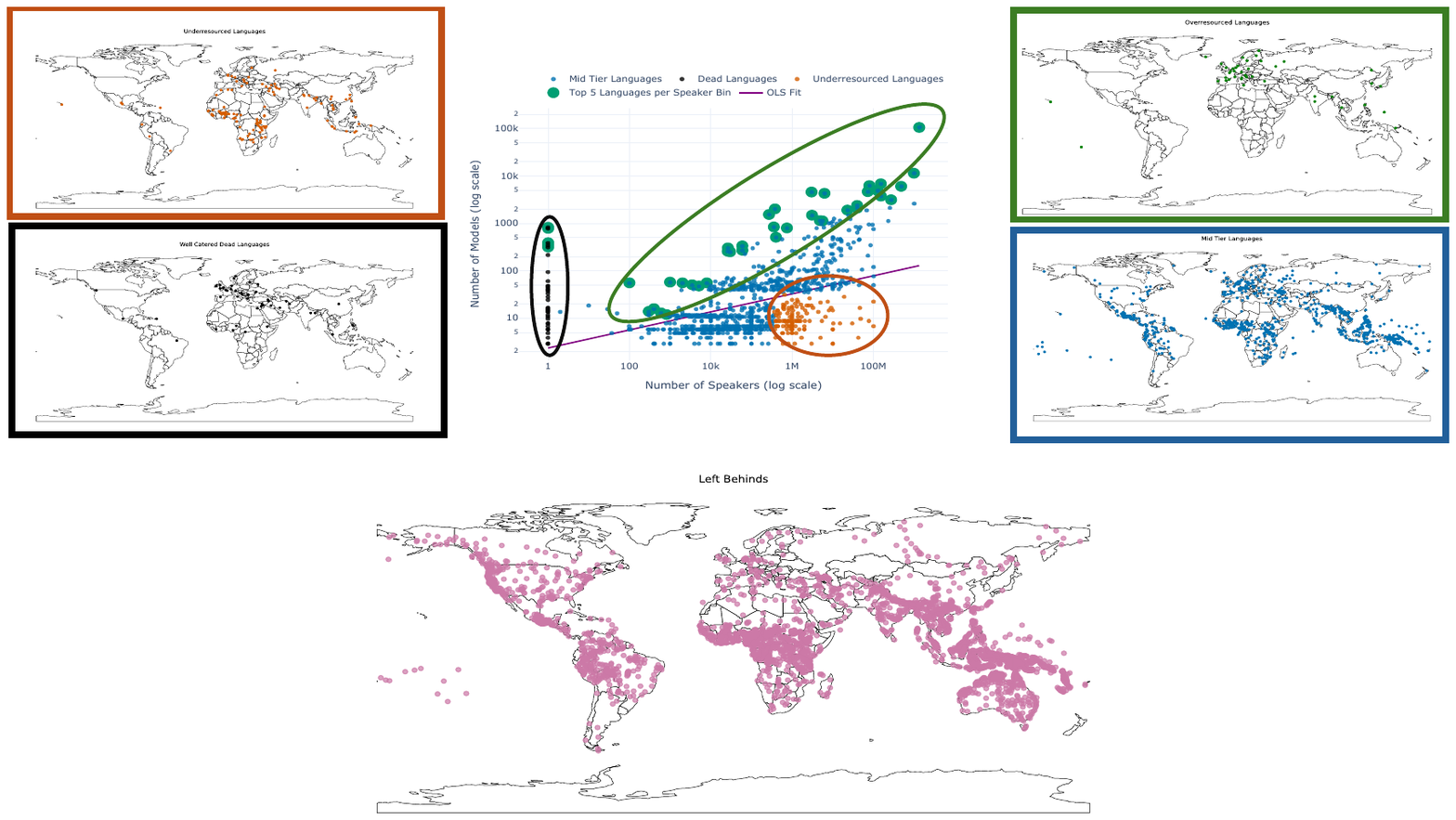}
    \caption{\textbf{Global Distribution of Languages by Language Models and Speaker Population}. We plot the relationship between number of speakers and number of online language models on a log-log scale. Languages are grouped into four categories: ``Mid-tier" (blue), ``Dead" (black), ``Under-resourced" (orange), and ``Top 5 Languages per Speaker Bin" (green). By “Top 5 Languages per Speaker Bin” we refer to the five languages with the most language models within each speaker population range. A purple OLS regression line indicates the expected trend (with parameters $\beta_1 = 0.312, \ p < 0.001$, $R^2 = 0.304$), with languages falling below it, despite having over 1 million speakers, classified as under-resourced. Each group is plotted on world maps showing their geographic distribution, with expanded subplots in larger size available in the Supplementary Figures in  \ref{geo_bigger}.}
    \label{fig:geographical_distr} %
\end{figure}

We then examined whether any geographical or demographic factors help explain the observed inequalities in language technologies. We first analysed the relationship between language model availability and speaker population size across our sample of languages. Using ordinary least squares regression, we modelled the logarithm of the number of language models (for languages with at least 1 model) against the logarithm of the speaker population, revealing a positive but weak correlation ($\beta_1 = 0.312, \ p < 0.001$, $R^2 = 0.304$); see Fig \ref{fig:geographical_distr}. This relationship is robust to the exclusion of the most-spoken languages: removing the top 5 to 20 languages by speaker population leaves the correlation moderate and highly significant (Pearson $r$ falling only from 0.55 to 0.51; Supplementary Table \ref{tab:sensitivity}). While this suggests that languages with larger speaker bases tend to have more models, notable deviations from the regression line expose systemic biases.

Languages falling significantly below predicted values, defined as having studentized residuals $p<0.05$ and speaker populations greater than 1 million, are concentrated in Sub-Saharan Africa, South Asia and the Middle East. For example, Nigerian Pidgin ($\sim$85 million speakers, 10 models, Studentized Residual = -1.5648, $p$ = 0.05), while Chittagonian ($\sim$13 million speakers) has only 1 model, Studentized Residual = -1.9672, $p$ = 0.02. Wu Chinese ($\sim$80 million speakers), despite its global prominence, has 10 models (Studentized Residual = -1.6126, $p$ = 0.05). These gaps persist even in technologically advanced regions: 2,171 languages without models include clusters in high-income areas such as California (USA) and Queensland (Australia), averaging 50.243 speakers (SD=522.405). The disparity is stark when contrasted with Breton (France, $\sim$200,000 speakers), which has as many models as Igbo (Nigeria, Cameroon, Equatorial Guinea, $\sim$31 million speakers). Such findings directly challenge recent industry pledges to “leave no language behind” \citep{costa2022no}.

In contrast, languages with disproportionately many models relative to their speaker populations are overwhelmingly European. Finland exemplifies this pattern, with Finnish, Ingrian, Inari Saami, North Saami, and South Saami among the least-spoken languages with the highest model counts. This reflects the European Union's institutionalised support for minority language preservation \citep{rehm2020european, rehm2023european}. Unexpectedly, several dead European languages like Latin, Ancient Greek, and Old English, also show elevated model counts, likely reflecting their academic prestige, copyright-free corpora, and cultural symbolism. Yet this investment does not align with contemporary needs, raising questions about prioritisation in language technology development. Looking at the languages left behind, it is clear that coverage gaps are not simply a Global North–Global South divide: in highly developed countries such as the United States and Australia, all languages other than English fall into this category. 

In the Supplementary Figure \ref{fig:combined_images_3}, we show this phenomenon by comparing the proportion of languages covered by AI in two countries generally regarded as highly developed,  Australia and the United States against Nigeria, a country with a different development trajectory and a rich linguistic landscape. This pattern is, in one sense, unsurprising: states allocate resources to particular languages in a manner consistent with pre-existing language policies, and these policies are themselves determined by factors such as official status, investment in education, and the historical dominance of specific language varieties, rather than by technological readiness alone. Nonetheless, this situation sits uneasily alongside commitments such as those articulated in the UN International Decade of Indigenous Languages (2022–2032), which affirms the rights of Indigenous peoples to preserve, revitalise, and promote their languages \citep{unLinguisticDiversity, UNESCO_OER_2019}. The states with the greatest technical capacity to implement these commitments are frequently among those least likely to be fulfilling them, a discrepancy that EQUATE renders visible at the level of individual languages rather than through national-level aggregates.

In sum, our results reveal a dual pattern: a persistent under-representation of minority languages, even in technologically advanced regions, and a persistent over-representation of European languages, including extinct ones, driven by historical and institutional factors. These findings highlight the need for transparent, needs-based frameworks to guide resource allocation and ensure equitable access to language technologies across diverse linguistic ecosystems; precisely the gap our Readiness Index is designed to address.

\newpage

\subsection{The uneven diffusion patterns of language technologies}

The unprecedented pace of conversational AI proliferation motivates an examination of its diffusion dynamics. To this end, we estimate the number of individuals worldwide who were covered by at least one ready-to-use conversational AI model between 2020 and 2024 (see Supplementary Materials in \ref{data}). Whereas the adoption of conventional technologies is typically quantified in terms of unit sales or per-capita ownership (as is standard for mobile phones or personal computers), most conversational agents constitute intangible assets within the digital economy and are openly accessible \citep{brynjolfsson2002intangible}. We therefore employ the number of speakers covered by at least one ready-to-use conversational AI model as a proxy measure for diffusion.

\begin{figure}[h!]
    \centering
    \includegraphics[width=\linewidth]{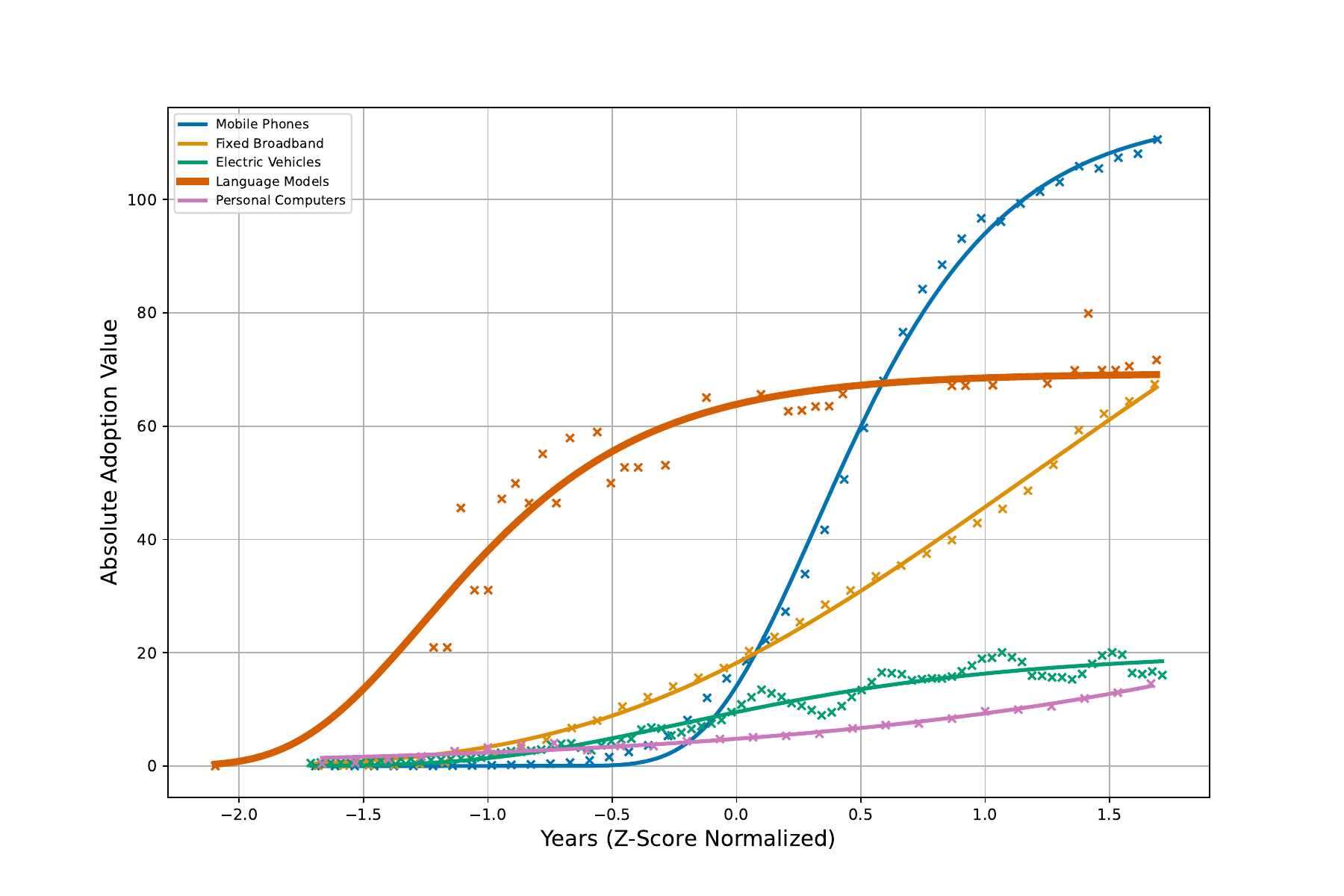}
    \caption{\textbf{Linearized Gompertz transformation of adoption data for Language Models, Mobile Phones, Fixed Broadband, Electric Vehicles and Personal Computers.} Growth for each technology progresses from the bottom-left (early adoption) towards the upper-right (approaching saturation). We normalise both adoption values (to maximum) and time (Z-score) to enable a cross-technology comparison. }
    \label{fig:comparative_gompertz}
\end{figure}

We discover that while most technologies follow an S-shaped adoption curves: low uptake, rapid growth, then saturation \citep{rogers2003diffusion}, language technologies diverge sharply. Fitting Gompertz curves (a sigmoid function which describes growth as being slowest at the start and end of a given time period) to longitudinal release data \citep{geroski2000models}, we find that mobile phones, PCs, and even electric vehicles adhere to classic S-shaped patterns. However, language models exhibit an earlier increase (displacement rate $b$ = 0.927; growth rate constant $c$ = 1.31, $R^2$ = 0.866), Fig \ref{fig:comparative_gompertz}, indicating ``hyper-growth": rapid acceleration at the very start of diffusion, without the slow initial-adoption phase seen in classical S-shaped curves, corresponding to a low Gompertz displacement (see Methods in Section \ref{stat_mod}). We report the parameters of the Gompertz fit for each technology in Supplementary Table \ref{fig:combined_gompz_plots_with_legend}). 

This divergence is due to structural differences. Unlike technologies that diffuse incrementally across communities, language models are trained in hundreds of languages simultaneously \citep{conneau2019cross}, obscuring representation gaps. Adoption lacks organic bottom-up dynamics; instead, inclusion is driven by top-down industrial priorities and global web data availability, reinforcing a “rich-get-richer' Zipfianization process. The observed deceleration in model adoption does not signal equitable catch-up. Rather, it reflects the consolidation of technological dominance, instigating inequalities. Early hyper-growth, driven by breakthroughs in dominant languages such as English, created steep advantages. As expansion slows, under-resourced languages struggle to close gaps, with development efforts focusing on refining high-performing models rather than broadening access. This two-stage dynamic (hyper-growth followed by lock-in) creates a self-reinforcing feedback loop, amplifying linguistic and socioeconomic disparities.

Our findings question the notion of the universal applicability of language models as General Purpose Technologies (GPTs) \citep{bresnahan1995general,brynjolfsson2017artificial,goldfarb2023could,eloundou2024gpts}. For a GPT to serve as a catalyst for widespread economic and social transformation, its advantages must be disseminated across various sectors and communities. Nonetheless, the predominance of technology in a limited number of dominant languages results in the exclusion of significant portions of the global population, thereby restricting their tangible effects. Paradoxically, the commercial promotion of these technologies has led to a reduction in investment in initiatives that focus on community engagement, as the multilingual solutions offered by major technology companies are deemed adequate \citep{Kreutzer, T20Brasil_TF05_2024}. Consequently, the development of language models may exacerbate rather than ameliorate existing disparities.

\section{The EQUATE AI Readiness Index: Closing the AI Language Divide}

We introduce the Language AI Readiness Index (EQUATE), an open-source metric designed to quantify both the urgency and the feasibility of language AI development for all attested languages. To construct this index, we assembled and harmonized data on existing language AI resources, the necessary digital infrastructure for AI deployment, and salient socioeconomic conditions, drawing on peer-reviewed research and data products issued by international organizations. These three domains were selected because prior work consistently identifies them as the principal determinants of communities’ capacity to develop, deploy, and sustain language technologies \citep{nigatu2024zeno}. The specific academic literature underlying each indicator is documented in the Appendix, with an extended review in Section~\ref{lit} and a concise summary in Table~\ref{tab:data_summary}.

It is important to underscore that AI readiness, as operationalized here, quantifies the feasibility of developing language AI for a given language; it does not address the normative question of whether such development ought to occur. A high readiness score does not entail that a community desires or endorses the introduction of these tools. Feasibility and demand are analytically distinct constructs, and we contend that demand can only be adequately established through direct, participatory engagement with speaker communities (e.g., \citet{mager2023ethical}).

The development of EQUATE proceeded in two stages. First, we conducted an in-depth correlational and dimensionality analysis of the collected indicators to uncover their underlying structure and interdependencies. This statistical analysis ensures that the variables included in the index are both representative and non-redundant, and that their grouping into broader dimensions reflects empirically observed relationships in the data. The methodological details of this analysis are provided in the Methods in Section ~\ref{stat_mod}.

Second, we used the results of this data-driven analysis to inform a global expert survey, which elicited the relative weights assigned to each indicator when computing the index. This step incorporates expert judgment from multilingual AI and technology communities, aligning the empirical structure of the data with the perspectives of practitioners actively engaged in language technology development. Languages are then categorised into readiness tiers based on their composite scores, which can be examined as a unified measure or along three decomposed dimensions: AI resources (datasets, models, tools), digital infrastructure (computational access, connectivity), and socioeconomic conditions (demographics, sociolinguistics and economic readiness). Details of the survey methodology are presented in Section~\ref{survey}, and respondent characteristics and results are reported in Supplementary Sections~\ref{survey_participants} and~\ref{surveyr_results}.

We start by examining the relationships among the collected features. The correlation matrix (Supplementary Figure \ref{fig:corr_matrix}) reveals two distinct clusters of features: the AI resources on one hand, and the socioeconomic and digital infrastructure resources on the other. This separation suggests that technological and social readiness evolve along partially independent axes. 

Principal component analysis (PCA) further supports this structure, revealing two dominant components that together account for 58.4\% of the total variance. The number of retained components was determined by inspecting the inflection point of the scree plot (elbow criterion), which indicated a clear drop in explained variance after the second component. After varimax rotation, the first component was dominated by social and digital infrastructure indicators --- including the proportion of individuals using the internet (loading = 0.35), Human Development Index (HDI; loading = 0.35), and education level (loading = 0.33 --- while the second captured the availability of AI technologies, with strongest loadings for Wikipedia active users (0.37), amount of GB on CommonCrawl (0.37), and number of datasets on HuggingFace (0.37) (see Supplementary Fig.~\ref{fig:pca_1}).

\begin{figure}[h!]
    \centering
    \includegraphics[width=\linewidth]{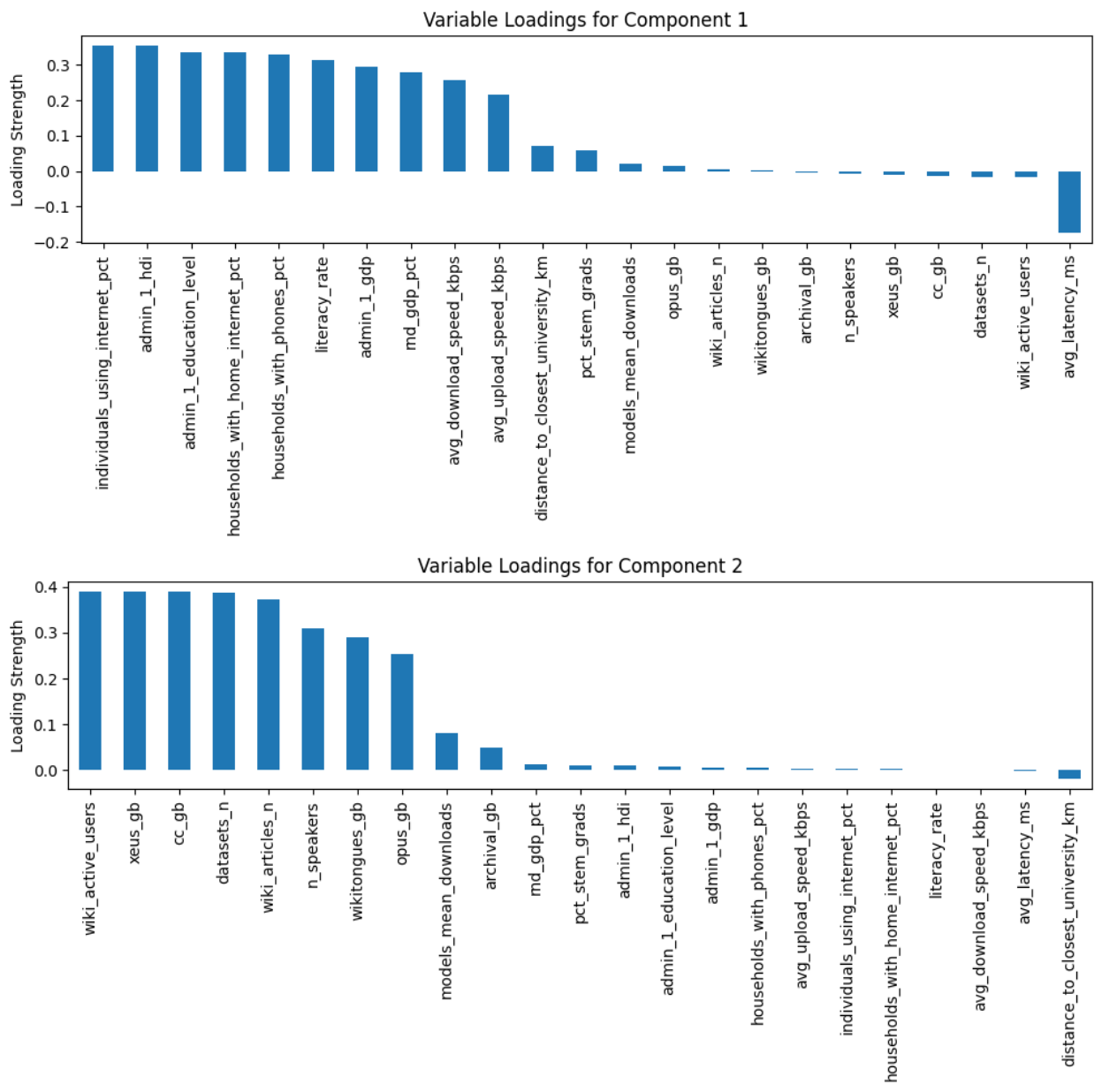}
    \caption{Bar plots displaying the magnitude and direction of the component loadings for 24 variables on the first two principal components (PC1 and PC2) after Varimax rotation. PC1 (top panel) shows a strong positive contribution from socioeconomic and digital infrastructure variables, with the highest loading strengths observed for the proportion of individuals using the internet (0.35), the Human Development Index (HDI; 0.35), and education level (0.33). Network latency provides the sole significant negative loading ($-0.18$). PC2 (bottom panel) is primarily defined by the availability of AI resources, with the highest loading strengths coming from Wikipedia active users (0.35), the number of datasets (0.35), and CommonCrawl data volume (0.35). The clear separation of variable groups across the two components graphically confirms the distinct factors identified by the PCA.}
\label{fig:pca}
\end{figure}

To assess the predictive relationships behind AI readiness, we set-up a regression analysis modeling the number of available AI models as the outcome variable. The number of datasets was excluded due to multicollinearity, and two binary variables (existence of a Bible and language institutional status) were included. A stepwise (forward–backward) linear regression within a mixed-effects framework (random effects: language family, $n=236$; primary country, $n=184$; macroarea, $n=6$) selected eight predictors, of which four AI-related features showed significant fixed effects: (existence of a Bible: $\beta=3.287$, $SE=0.358$, $t(5.117)=9.170$, $p<0.0001$; number of speakers: $\beta=1.690$, $SE=0.148$, $t(4.920)=11.438$, $p<0.0001$; GBs of information in the OPUS corpus: $\beta=0.974$, $SE=0.202$, $t(5.426)=4.824$, $p<0.0001$; GBs of information in the XEUS corpus: $\beta=1.232$, $SE=0.151$, $t(5.252)=8.182$, $p<0.0001$) (Fig \ref{fig:pca}).

When the two PCA-derived components were used instead, both significantly predicted the number of models (AI resources PC: $\beta=0.229$, $SE=0.014$, $t(2412.337)=16.725$, $p<0.001$; socioeconomic/digital infrastructure PC: $\beta=0.215$, $SE=0.044$, $t(269.625)=4.915$, $p<0.001$). The PCA-based model provided a substantially better fit (BIC = 15,180.98) compared with the stepwise regression (BIC = 31,648.98), indicating that technological and social readiness contribute complementary and independent effects to model development (Supplementary Figure \ref{fig:stepwise}). 

Taken together, these findings substantiate the conceptual design of the EQUATE index by empirically demonstrating that its AI resource, socioeconomic, and digital infrastructure dimensions are both statistically distinct and jointly predictive of language AI development. That is, we provide quantitative evidence that readiness should be conceptualized as a multidimensional construct, determined not only by data availability but also by the broader social and infrastructural conditions that enable or constrain the emergence of language technologies. Although prior work has made this article qualitatively through case studies \citep{nigatu2024zeno} or proxied demand primarily through speaker population size \citep{joshi2020state, blasi-etal-2022-systematic}, to our knowledge, this study is the first to establish language AI readiness as a multidimensional construct through large-scale empirical investigation.

\begin{figure}[h!]
    \centering
    \includegraphics[width=\linewidth]{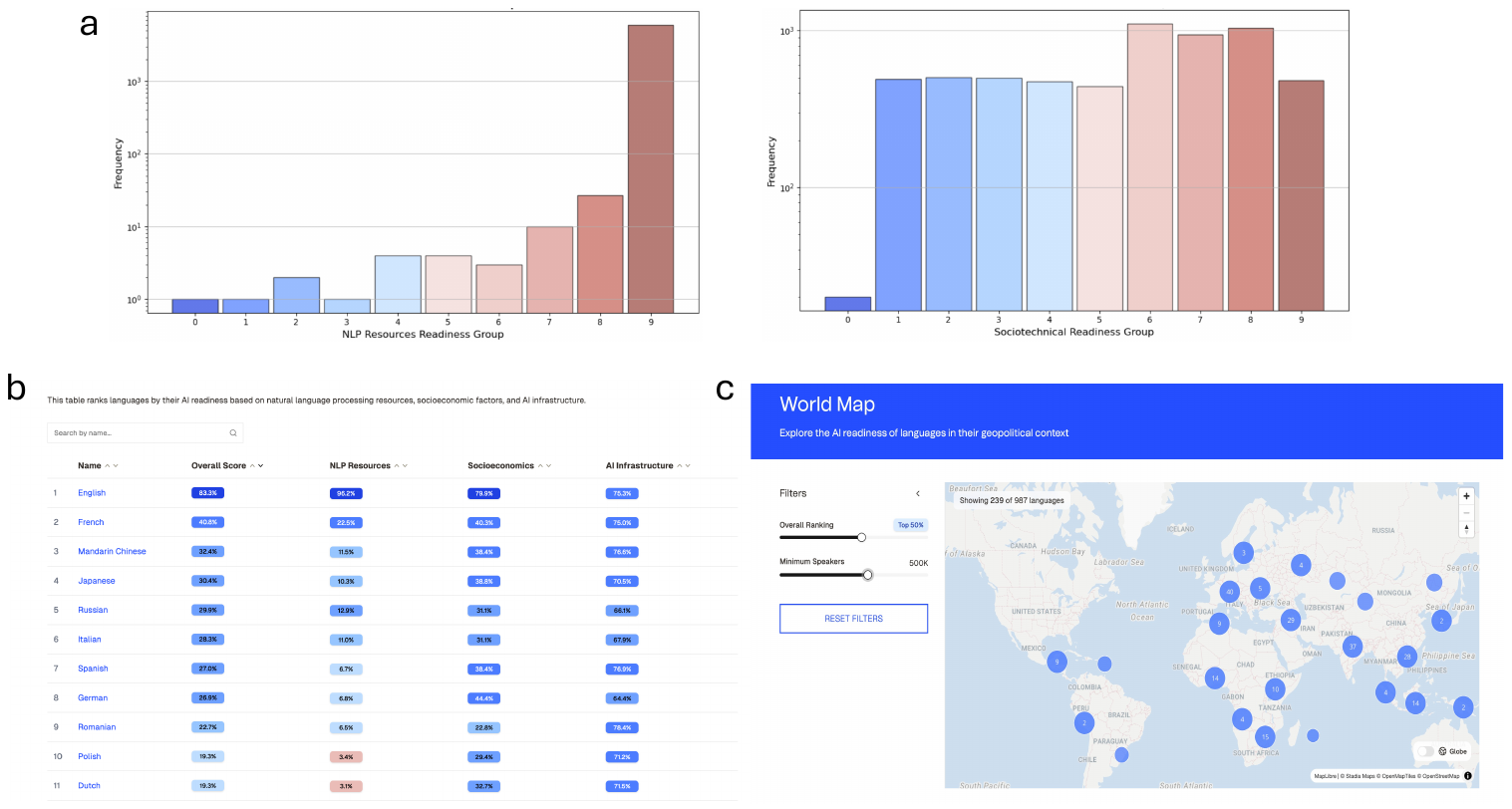}
    \caption{\textbf{Demonstration of our open access AI language readiness tool.} In subfigure \textbf{a}, we show the differential in distribution between languages having AI resources and languages being socio-technically ready for AI. We observe how while many languages might have a fraction of the language modeling resources available compared to English, they still do have the readiness in terms of socioeconomics and digital infrastructure. Below, in subfigure \textbf{b} we show the top ranking languages in our index ranked by overall score, available at \url{https://www.equate-index.ai/}. In subfigure \textbf{c} we show the global explorer feature of our index. Blue circles indicate language clusters, with the number inside showing the count of languages in each region. The markers are interactive, allowing zoom and click to reveal individual languages and their AI readiness metrics. Filters on the left enable customization by overall ranking and minimum speaker count. 
}
    \label{fig:index}
\end{figure}

\pagebreak

\section {Discussion}

In this study, we introduce an open-source instrument called the EQUATE index, designed to assess both the urgency and the feasibility of advancing language AI technologies across all attested languages. To construct this index, we systematically compiled and harmonized data on existing language AI resources, the necessary digital infrastructure for AI deployment, and salient socioeconomic indicators, drawing on peer-reviewed scholarship and data products disseminated by international organizations. As language technologies become increasingly integrated into everyday practices, the unprecedented disparities we document warrant carefully targeted interventions to mitigate the risk of exacerbating existing social and economic inequalities. 

We demonstrate that the EQUATE index translates the above insight into a single, interpretable, language-specific metric. Contrary to the widespread assumption that most of the world's languages are technologically unprepared, our results show that more than a thousand languages, beyond those already well served by AI, possess the foundations needed to develop language technology. For many of these, the foundation is conventional data: web text, parallel corpora, and Bible translations. For others, it lies in less standard but still substantial resources, including grammatical descriptions, documentary-linguistics archives, and community-led language-preservation initiatives. This holds even when Bible translations are excluded. Among these, 597 languages—predominantly in Africa and Asia—currently have fewer than three language technologies or models, yet nonetheless display characteristics indicative of AI readiness.

In addition, we identify languages that, despite lacking a substantial digital footprint, satisfy key socioeconomic and digital infrastructure prerequisites for AI integration. Examples include Hakka Chinese (with approximately 80 million speakers globally) and Darija (Moroccan Arabic, spoken by more than 90\% of Morocco’s approximately 40 million inhabitants). In several cases, this misalignment between infrastructural readiness and technological provision appears to stem from a systematic preference for developing technologies for standardized or “prestigious” varieties—such as Mandarin or Modern Standard Arabic—on the grounds of their perceived broader applicability. Such prioritization, however, may have detrimental consequences for non-standardized local varieties, thereby deepening the digital marginalization of already underserved communities.

Conversely, we observe that languages such as Esperanto and Interlingua, constructed languages with approximately zero native speakers, receive a disproportionately high level of attention from the AI research community. This pattern exemplifies a broader misalignment between research priorities, the personal interests of researchers, and the actual linguistic needs of speaker communities. Factors such as the availability of clean, well-structured datasets, longstanding academic traditions, and the convenience of using these languages for benchmarking frequently take precedence over considerations of real-world applicability, thereby impeding progress toward linguistic equity in AI development.

Our open-access tool, reproduced in Fig \ref{fig:index}, enables users to systematically rank languages according to their AI readiness scores, both as aggregate measures and disaggregated by individual dimensions. It further supports the examination of AI readiness for languages spoken across different national contexts and the analysis of how languages compare within their broader socioeconomic environments. EQUATE allows users to rank languages based on global or country-specific readiness scores and to identify regional disparities. 

The tool provides a methodological framework for promoting more equitable AI development that is aligned with community needs and constrained by technological feasibility. By mapping the structural conditions that determine the feasibility of AI deployment, the index identifies contexts in which targeted interventions may counteract cumulative advantage dynamics. It particularly emphasizes languages whose technical, data, or infrastructural capacity exceeds their current level of AI support—languages with “underutilized potential” that constitute strategic leverage points for investment and community-based collaborations. To ensure the index remains current and policy-relevant, we will update it annually with new data sources and indicators, and we will maintain an open-access platform to facilitate contributions from AI developers, policymakers, language communities, and other stakeholders.

The index offers actionable guidance to researchers, developers, investors, policymakers and other stakeholders for strategic resource allocation across diverse linguistic contexts. For example, developers can use it to identify under-served languages with sufficient data availability and work with speaker communities to advance AI development. For investors, it highlights settings with strong enabling conditions but limited language AI presence, where targeted support could substantially accelerate progress. Policymakers can draw on the index to identify contexts where foundational infrastructure or socioeconomic conditions require strengthening to enable sustainable language technology. Language communities can use it as a basis for understanding where their language stands and for advocating for the resources and tools they choose to prioritize. Across all settings, close collaboration with local stakeholders remains essential to ensure culturally relevant and impactful solutions and to address persistent barriers such as digital literacy gaps.

Future research must deepen our understanding of the societal needs of speaker communities, spanning their social, cultural, and economic contexts, and identify the types of applications that can meaningfully support progress in these areas. Because most populations use multiple languages for different purposes, languages should be examined within their multilingual ecosystems rather than in isolation \citep{bird-2022-local}. Technologies designed for cultural transmission, for example, require capabilities distinct from those needed in technical or administrative domains. Grounding future research in these local priorities and real-world language practices will be essential for shaping language AI that supports sustainable and equitable development.

Nonetheless, our study presents a framework for identifying global disparities in technological readiness across languages and communities. Combining evidence on AI resources, digital infrastructure, and socioeconomic conditions, the index offers data-driven guidance for language AI stakeholders. It helps identify priority areas and supports targeted investments and policies that reduce, rather than reinforce, digital divides in language AI development.

\section{Methods}

\subsection*{Statistical models and basic data collection}\label{stat_mod}

First, we collected a list of languages having at least one documented vocabulary item and some level of geographical information, finding a total of n = 6003 languages. We select only attested languages, meaning those languages that are documented --- living or extinct --- with surviving evidence such as texts, recordings, or inscriptions. By contrast, unattested languages lack direct evidence and may be known only by name, lost entirely, or reconstructed as proto-languages.
We extract language information and geographical centroids of languages from \url{glottolog.org} \citep{nordhoff2011glottolog}, an open-access online bibliographic database of the world's languages. The geographical centroids are crucial to map languages to the relevant sub-national socioeconomic statistics; hence, languages without a documented geographical location are excluded from the dataset. The existence of vocabulary items for languages not covered by a Bible or by scriptures is attested through consulting PanLex \citep{kamholz2014panlex}, a lemmatic translation resource that combines a large number of translation dictionaries and lexical resources, as well as multiple archives of linguistic documentation reported in the Supplementary Data information \ref{data}. We tested our hypotheses through the following models: 

\begin{itemize}
    \item 

    H1: Although the number of languages included in AI research is increasing, technological advancements disproportionately benefit languages that are already well-resourced, resulting in a `rich-get-richer' dynamic.
\end{itemize}
We run a linear regression analysis on the log-transformed data. By transforming the power-law relationship $x(k) = \frac{C}{k^\alpha}$ into $\ln(x(k)) = \ln(C) - \alpha \ln(k)$, we obtain a linear model where $\ln(x(k))$ is the dependent variable ($Y$), $\ln(k)$ is the independent variable ($X$), $-\alpha$ is the slope, and $\ln(C)$ is the intercept. Hypothesis H1 is supported if the estimated slope $(-\alpha)$ is negative and statistically significant, indicating a power-law distribution in which a small number of languages (the ``richer" ones) capture a disproportionately large share of language technology development, represented by $x(k)$.

For models, we fit a two-segment (broken) power law, motivated by a visible change in slope in the head of the distribution. We fit two linear regressions on the log-transformed data, partitioned at a breakpoint rank $k^{*}$: a head regime over ranks $k \le k^{*}$ and a tail regime over ranks $k > k^{*}$,
\begin{align}
  \ln(x(k)) &= \ln(C_1) - \alpha_1 \ln(k), \quad && k \le k^{*}, \\
  \ln(x(k)) &= \ln(C_2) - \alpha_2 \ln(k), \quad && k > k^{*},
\end{align}
where $\alpha_1$ and $\alpha_2$ are the head and tail exponents and $\ln(C_1), \ln(C_2)$ the corresponding intercepts. The breakpoint $k^{*}$ is selected by minimising the combined residual sum of squares of the two segments over all admissible split ranks, subject to a minimum segment size, and the resulting two-segment model is compared against the single power law by the Bayesian Information Criterion (BIC) to confirm that the additional segment is
warranted.

The crossover rank $k_{c}$ is the rank at which the two fitted regime lines intersect, obtained by equating the two log-linear models,
\begin{equation}
  \ln(C_1) - \alpha_1 \ln(k_c) = \ln(C_2) - \alpha_2 \ln(k_c),
\end{equation}
which solves to
\begin{equation}
  \ln(k_c) = \frac{\ln(C_1) - \ln(C_2)}{\alpha_1 - \alpha_2},
  \qquad
  k_c = \exp\!\left(\frac{\ln(C_1) - \ln(C_2)}{\alpha_1 - \alpha_2}\right).
\end{equation}

\begin{itemize}

\item H2: We hypothesize that language technology development does not follow traditional S-shaped adoption curves, but is dictated by rapid forces in the development of AI 
\end{itemize}
To analyze the diffusion of language technologies, we fit a Gompertz model, which captures the typical S-shaped growth pattern of technology adoption, including slow initial uptake, rapid acceleration, and eventual saturation. The primary metric for cumulative adoption is the total count of speakers covered, $S_t$, calculated at each time step $t$ as

$$
S_t = \sum_{L \in \mathcal{L}} P_L \cdot M_{L,t},
$$

where $P_L$ is the number of speakers of language $L$, $\mathcal{L}$ is the set of all languages considered, and $M_{L,t}$ is a binary indicator defined as

$$
M_{L,t} = \begin{cases} 
1 & \text{if language } L \text{ has at least 1 openly available conversational language models at time } t \\ 
0 & \text{otherwise} 
\end{cases}.
$$

This formulation accounts for multilingualism by summing speakers across all languages meeting the coverage threshold, thus measuring speaker coverage rather than distinct individuals. The empirical $S_t$ series is then used to fit the Gompertz function

$$
L(t) = A e^{-B e^{-Ct}},
$$

where $A$ is the upper asymptote (maximum potential speaker coverage), $B$ is a displacement parameter controlling horizontal position and inflection timing, $C$ is the growth rate, and $t$ is time. Parameters $A$, $B$, and $C$ are estimated via non-linear regression to minimise the sum of squared differences between observed $S_t$ and predicted $L(t)$.

To contextualise the diffusion dynamics of language technologies, we compare their fitted Gompertz curves with those of other established technologies. For each dataset, we standardized the independent variable, time ($t$), via Z-score normalization. This process, defined as $t_{norm} = \frac{t - \mu_t}{\sigma_t}$, where $\mu_t$ and $\sigma_t$ are the mean and standard deviation of the time series, respectively, is a crucial step to ensure the numerical stability of the non-linear least squares regression. By centering the time variable at zero with a unit standard deviation, we allow the comparability of growth rate across technologies with different historical starting points and durations.

\subsubsection*{Index data analysis – Stepwise and PCA-based regression}\label{stat_mod}

To identify features predictive of the number of AI models, we implemented two complementary approaches: a stepwise hierarchical linear mixed-effects regression and a PCA-based model.

For the stepwise regression, we fitted a hierarchical linear mixed-effects model of the form:

\[
Y_{ijk} = \beta_0 + \sum_{m=1}^{M} \beta_m X_{m,ijk} 
+ u_i + v_{ij} + w_k 
+ \sum_{m=1}^{M} u_{m,i} X_{m,ijk} + \sum_{m=1}^{M} v_{m,ij} X_{m,ijk} + \sum_{m=1}^{M} w_{m,k} X_{m,ijk} 
+ \epsilon_{ijk}
\]

where $Y_{ijk}$ is the log-transformed number of models for language $k$ in country $j$ and macroarea $i$; $X_m$ represents the fixed-effect predictors; $u_i$, $v_{ij}$, and $w_k$ are random intercepts for macroarea, primary country nested within macroarea, and language family, respectively; $u_{m,i}$, $v_{m,ij}$, and $w_{m,k}$ are corresponding random slopes; and $\epsilon_{ijk}$ is the residual error.

Stepwise selection combined forward and backward procedures, iteratively adding or removing variables to optimize model fit according to the Bayesian Information Criterion (BIC). At each step, all candidate variables not yet included (forward) or included (backward) were evaluated, and the variable whose addition or removal most improved the model fit was selected. The process terminated when no further improvement was possible.

For the PCA-based model, all continuous features were standardized and subjected to principal component analysis (PCA). Two components were retained based on the scree plot elbow criterion, followed by varimax rotation. The resulting component scores were used as predictors in a hierarchical linear mixed-effects model with random intercepts and random slopes, specified as:

\[
Y_{ijk} = \beta_0 + \sum_{p=1}^{P} \beta_p PC_{p,ijk} 
+ u_i + v_{ij} + w_k 
+ \sum_{p=1}^{P} u_{p,i} PC_{p,ijk} + \sum_{p=1}^{P} v_{p,ij} PC_{p,ijk} + \sum_{p=1}^{P} w_{p,k} PC_{p,ijk} 
+ \epsilon_{ijk}
\]

where $PC_p$ represents the score of the $p$-th principal component, and the random intercepts and slopes are defined analogously to the stepwise model.

\subsection*{Reanalysis of existing datasets}
\subsubsection*{The ACL anthology} \label{acl_method}

Here we analyse data presented in \citet{radev2013acl}. From the dataset, we identify a total of n = 18183 papers matching a language, covering 328 languages. 

We move beyond the ACL Anthology in our analysis for two main reasons. First, many language technology applications are developed outside of academic publishing. Second, published papers often fail to specify the languages they target, obscuring a clear picture of which languages underpin AI applications \citep{bender2019benderrule}. 

\subsection*{Index computation}

Below we present the  Index computation. The total index will be decomposed into 3 indices: an index of AI resources readiness, an index of socioeconomic readiness, and an index of digital infrastructure readiness. 
\begin{enumerate}

    \item \textbf{Merging Correlated Features} \\
    When two features $f_1, f_2 \in F_g$ show high linear dependence (e.g., Pearson correlation $> 0.85$), they are combined into a single composite feature to reduce redundancy:
    \begin{align*}
        x_{i, f_{\text{merged}}} &= \frac{w_{f_1} x_{i,f_1} + w_{f_2} x_{i,f_2}}{w_{f_1} + w_{f_2}}, \\
        w_{f_{\text{merged}}} &= w_{f_1} + w_{f_2}
    \end{align*}

    \item \textbf{Shifting continuous features} \\
    To prevent zero-valued terms in geometric aggregation, a value of 1 is added to all non-binary numeric features:
    \[ x_{i,f} \leftarrow x_{i,f} + 1, \quad \text{for } f \notin B \]

    \item \textbf{Log-Transform (Optional)} \\
        Several count- and size-valued features are extremely right-skewed, which
    min--max normalisation alone would compress towards zero. We therefore screen
    features by sample skewness and log-transform those with high positive skew. For features selected for logarithmic scaling:
    \[ x_{i,f} \leftarrow \log(x_{i,f} + \varepsilon) \]
    This transformation is useful when a feature shows exponential characteristics, often revealed through visual or statistical diagnostics.
    \begin{itemize}
        \item \textit{Note:} Since we already added 1 in Step 2, $x_{i,f}$ will be $\ge 1$. The $\varepsilon$ is a general safeguard for the logarithm function, ensuring the argument is strictly positive, though in this sequence it's less critical.
    \end{itemize}

    \item \textbf{Normalization of continuous features} \\
    To ensure all non-binary features contribute equitably regardless of their original scale, and after any optional log-transformation, they are normalized to a common range, typically $[0, 1]$. For a feature $f \notin B$:
    \[ x_{i,f} \leftarrow \frac{x_{i,f} - \min_j(x_{j,f})}{\max_j(x_{j,f}) - \min_j(x_{j,f})} \]
    To prevent zero values in subsequent geometric aggregation (if the minimum value becomes 0 after normalization),  A deliberate lower bound $\delta$ is then imposed,
    \[
        x_{i,f} \leftarrow \max\!\left(x_{i,f},\, \delta\right),
        \qquad \delta = 0.01,
    \]
    where $\delta$ is a modelling choice (distinct from the numerical safeguard
    $\varepsilon$) that limits how severely a bottom-of-distribution feature is
    penalised in the geometric mean. If a feature is constant across all
    languages ($\max_j x_{j,f} = \min_j x_{j,f}$) it carries no discriminatory
    information and is assigned the neutral value $0.5$.

    \item \textbf{Geometric aggregation by group} \\
    For each group $g$, features are combined using a weighted geometric mean.
    \begin{itemize}
        \item \textbf{Case 1: renormalized weights} \\
        When $\sum_{f \in F_g} w_f \neq 1$, normalize:
        \begin{align*}
            \tilde{w}_f &= \frac{w_f}{\sum_{f' \in F_g} w_{f'}}, \\
            G_{i,g} &= \prod_{f \in F_g} x_{i,f}^{\tilde{w}_f}
        \end{align*}

        \item \textbf{Case 2: already normalized} \\
        If weights sum to 1:
        \[ G_{i,g} = \prod_{f \in F_g} x_{i,f}^{w_f} \]
    \end{itemize}

    \item \textbf{Aggregation across feature groups} \\
    A composite score across all groups is obtained as:
    \[ S_i^{\text{groups}} = \left( \prod_{g=1}^{G} G_{i,g} \right)^{1/G} \]

    \item \textbf{Binary feature adjustment} \\
    Each binary feature $b \in B$ imposes a multiplicative penalty when absent:
    \[ P_{i,b} = 1 - w_b (1 - x_{i,b}) =
        \begin{cases}
            1, & x_{i,b} = 1 \\
            1 - w_b, & x_{i,b} = 0
        \end{cases}
    \]
    Cumulative binary penalty:
    \[ P_i^{\text{binary}} = \prod_{b \in B} P_{i,b} \]

    \item \textbf{Final index score} \\
    The final readiness score for language $i$ combines continuous and binary components:
    \[ S_i = S_i^{\text{groups}} \cdot P_i^{\text{binary}} =
        \left( \prod_{g=1}^{G} G_{i,g} \right)^{1/G}
        \cdot
        \prod_{b \in B} \left( 1 - w_b (1 - x_{i,b}) \right)
    \]
\end{enumerate}
\subsection*{Weights determination through surveys}\label{survey}

To determine the weights for the index, we conducted two complementary expert surveys. Each targeted a distinct expert community and elicited rankings for a distinct set of factors. The first surveyed experts in multilingual language technology, who ranked the importance of AI-specific resources; the second surveyed experts in human–computer interaction and global technology adoption, who ranked digital-infrastructure and socioeconomic indicators. Participants were recruited from universities, international organizations, and industrial research labs. Selection prioritised two areas of expertise: multilingual language technologies and the sociotechnical dimensions of human-computer interaction and global technology adoption. 

\begin{itemize}
    \item \textbf{Data collection procedure}. Participants were recruited via email based on their demonstrated expertise in relevant domains, including multilingual AI, human-computer interaction, and global technology adoption. Expert selection was guided by publication records, conference participation (e.g., ACL, CHI), and institutional affiliations in research centers focused on language technologies or digital inclusion. Participants were invited to complete a Microsoft Forms survey in which they ranked the factors they considered most relevant for assessing a language's readiness for language technology development. Experts in multilingual AI ranked the importance of AI-specific resources, while experts in human-computer interaction and technology adoption ranked socioeconomic and digital infrastructure indicators.

    \item \textbf{Procedure}. Participants opened the Microsoft Forms URL and first provided their position, area of expertise, and institutional affiliation. They then completed the feature ranking task. Finally, they were given the opportunity to suggest additional features influencing language technology readiness.

    \item \textbf{Sample size} We recruited 30 experts from each field, with an emphasis on geographical diversity.
    \end{itemize}

\newpage
\clearpage
\bibliography{sn-bibliography}%
\newpage

\begin{appendices}

\section{Data availability statement}
\subsection{Existing data}\label{data}
We extracted data from publicly available online sources. All datasets are open access and are released alongside the study. Below, we list the datasets collected in relation to each research question.
\begin{enumerate}
    \item 
To assess the scale of inequality and its evolution, we collect monthly snapshots of HuggingFace (\url{https://huggingface.co/languages}), the main global platform for language models, using the Wayback Machine archive: \url{https://web.archive.org/web/20250000000000*/https://huggingface.co/languages}. We purposefully do not exclude multimodal artifacts, including for instance speech technologies such automatic speech recognition (ASR), text-to-speech (TTS). This is because language technologies should equally serve oral languages and speakers who do not write, and omitting speech would narrow the very long tail this study sets out to characterize.
\item To assess benchmark coverage per language, and to test whether the concentration observed in model counts persists under a stricter, evaluation-based measure, we recover benchmark availability from the primary literature. Benchmarks are collected from the ACL Anthology corpus \url{https://aclanthology.org/}, the main open-access repository of research in computational linguistics and natural language processing. It hosts the proceedings of the field's principal conferences and journals, published under the auspices of the Association for Computational Linguistics (ACL), and is the standard record of published work in the field. Benchmarks are collected from this corpus as follows.

The corpus is based on an ACL Anthology metadata snapshot downloaded on May 2026, covering all available venues from 2013 to the 2026 records present at that date. We first performed title-only screening to identify likely dataset, resource, and benchmark papers, using the query:
\begin{quote}
\small\ttfamily
("dataset" OR "datasets" OR "benchmark" OR "benchmarks" OR "corpus" OR "corpora" OR "shared task" OR "shared-task")
\end{quote}
This produced 7,417 title-screened papers. For each, the PDF was downloaded and text extracted with \texttt{pdfplumber} from the first 20 pages. The resulting paper-level corpus combines metadata, extracted page text, and extracted tables; figures are not processed.

Language matching was performed against the 6,003 language entries in the EQUATE index, each with a language name, ISO code, and Glottocode. For each paper we searched the title, abstract, extracted page text, and extracted table content. Language names were matched case-insensitively as complete token sequences using Unicode word tokenisation: multi-word names were counted only when the full token sequence appeared contiguously, and substring matches were disallowed (so that, for example, ``Ewe'' does not match inside ``fewer'').

To control precision, matches were separated into \emph{confirmed} and \emph{candidate} mentions. Multi-word language names were treated as confirmed, as was a manually retained list of common single-token names (English, French, German, Spanish, Arabic, Chinese, Hindi, Swahili, Tamil, Turkish, Vietnamese, and others). Remaining single-token names were resolved contextually: confirmed when accompanied by explicit language evidence (``Afar language'', ``Languages: Afar'', or a table with a language-related header), and otherwise retained only as candidate mentions where the surface form was plausibly a language but the context ambiguous (e.g.\ Afar is also the name of a people and a region). Candidate mentions were excluded from the paper-level language counts, so the reported coverage is a conservative lower bound. ISO codes were matched conservatively: a candidate code had to be an independent alphabetic token of length 2--4, delimited on both sides by punctuation or table-like separators (e.g.\ \texttt{(eng)}, \texttt{[eng]}, \texttt{| eng |}, \texttt{: eng ;}), and a confirmed match additionally required a strong language cue (\emph{language}, \emph{languages}, \emph{ISO}, \emph{code}, \emph{locale}, or \emph{lang}) within an 80-character window.

We acknowledge that this procedure is intrinsically conservative in its coverage: it includes only benchmarks documented in the ACL Anthology and therefore excludes evaluation work disseminated through other venues or remaining unpublished. Because the methodological design systematically prioritizes precision over recall at each stage, the resulting per-language counts should be interpreted as a strict lower bound on benchmark coverage.

\item Data for the diffusion of language technologies is computed by summing the number of speakers for each language covered by at least one openly available commercial large language model. We collect data from the following coversational models: 
\begin{itemize}
    \item Qwen 2.5 \url{https://chat.qwen.ai/}
    \item Claude 3.7 \url{https://claude.ai/}
    \item GPT-4.5 \url{https://chatgpt.com/}
    \item Grok 3 \url{https://grok.com/}
    \item Llama 4 (Meta AI)
    \item DeepSeek-R1 \url{https://chat.deepseek.com/}
    \item Mistral Large \url{https://chat.mistral.ai/chat}
    \item Gemini 2.0 \url{https://gemini.google.com/}
\end{itemize}
To compare the diffusion of language technologies with other technologies, we collect data from the International Telecommunication Union (ITU) and historical records on technology adoption from the World Bank:
\begin{itemize}
    \item Mobile Phones, ITU: \url{https://datahub.itu.int/data/?e=701&c=&i=178}
    \item Fixed Broadband Coverage, ITU: \url{https://datahub.itu.int/data/?e=701&c=&i=19303}
    \item Electric Vehicles: \url{https://newautomotive.org/global-ev-tracker}
    \item Personal Computers per 100 People, World Bank:  \url{https://www.econstats.com/wdi/wdiv_597.htm}
\end{itemize}

\item For the index, we collect three types of data: AI resources, socioeconomic indicators, and digital infrastructure metrics. Depending on availability, data is gathered at three geographic levels: language population, first subnational unit, and country. Most national-level data comes from the World Bank and the ITU. The dataset covers 6,003 languages across 217 countries and territories. Language geographical centroids are collected from \url{https://glottolog.org/}.
\begin{itemize}
    \item \textbf{AI resources} 
    \begin{itemize}
        \item Number of Speakers: from Wikidata (\url{https://www.wikidata.org/wiki/Help:Wikimedia_language_codes/lists/all})
        \item Bible Translations \url{https://www.bible.com/languages} and \url{https://scriptureearth.org/00eng.php}
        \item Number of Models: from HuggingFace (\url{https://huggingface.co/languages})
        \item Available Data (in bytes): from CommonCrawl (\url{https://commoncrawl.github.io/cc-crawl-statistics/}), OPUS (\url{https://opus.nlpl.eu/}), and Wikipedia
        \item Linguistic Documentation Archives:
        \begin{itemize}
            \item AILLA: \url{https://www.ailla.utexas.org/}
            \item ANLA: \url{https://www.uaf.edu/anla/}
            \item ELAR: \url{https://www.elararchive.org/}
            \item DOBES: \url{https://dobes.mpi.nl/}
            \item Pangloss: \url{https://pangloss.cnrs.fr/}
            \item Kaipuleohone: \url{https://scholarspace.manoa.hawaii.edu/communities/}
            \item PARADISEC: \url{https://www.paradisec.org.au/}
            \item Wikitongues: \url{https://wikitongues.org/archive/}
        \end{itemize}
    \item Availability of Keyboards: from Keyman \url{https://keyman.com/en/}. We consider the availability of digital keyboards (as opposed to physical keyboards).
    \item Level of Grammatical Documentation: from Glottolog \url{https://glottolog.org/meta/glossary#sec-descriptivestatusofalanguage}
    \end{itemize}

    \item \textbf{Socioeconomic data}
    \begin{itemize}
        \item umber of Speakers: from \url{https://www.wikidata.org}. Each language is counted by its total speaker population (First Language, L1 and Second Language, L2).We did not restrict counts to L1 speakers alone, as comprehensive L1-only estimates are not available at scale. Moreover, the underlying construct of L1+L2 is often problematic in itself: much of the world's multilingual poopulation frequently acquires more than one language in early childhood, meaning that assigning an individual to a single first language is often an arbitrary choice. 
        \item Distance to Nearest University (km): from \url{https://www.4icu.org/reviews/index2.htm}, calculated from language centroids using Euclidean distance:
        \[
        d_{ij} = \sqrt{(x_i - x_j)^2 + (y_i - y_j)^2}
        \]
        \item Subnational HDI: life expectancy, education, and income (\url{https://globaldatalab.org/shdi/}) \citep{smits2019subnational}
        \item Subnational GDP per capita: from \url{https://globaldatalab.org/shdi/} \citep{kummu2018gridded}
        \item Educational Attainment Score: from \url{https://globaldatalab.org/shdi/}
        \item Cybersecurity Law (binary indicator): from ITU \url{https://datahub.itu.int/data/?i=100103&s=8428}
        \item R\&D Expenditure (\% of GDP): \url{https://data.worldbank.org/indicator/GB.XPD.RSDV.GD.ZS}
        \item Literacy Rate and \% of STEM Graduates: various World Bank sources \url{https://databank.worldbank.org/US-STEM-(ISCED-and-Tertiary)/id/cd77ac48}
        \item Number of Active Wikipedia Users: from Wikidata, \url{https://www.wikidata.org/wiki/Special:ActiveUsers}. By `Active User', Wikidata refers to users (excluding bots) who have completed at least one action within the platform in the past 30 days. 
        \item Language Ecology: from Ethnologue and APiCS (Atlast of Pidgin and Creole language Structures, \url{https://apics-online.info/}). This indicator grounds Bird's (2022) ecology of languages \citep{bird-2022-local} in the Expanded Graded Intergenerational Disruption Scale (EGIDS) \citep{lewis2010assessing}, derived from Ethnologue. Each language is assigned to one of three types. Languages of wider communication, used across many countries, are classified as \texttt{standardized\_bridge}; institutional or developing languages (EGIDS levels 0--5) as \texttt{local\_standardized}; and all remaining languages as \texttt{local}. The assignment combines the EGIDS level with two attributes already present in the index, namely whether the language holds official status and the number of countries in which it is used. This yields 4,772 \texttt{local}, 1,165 \texttt{local\_standardized}, and 66 \texttt{standardized\_bridge} languages. In line with Ethnologue's terms of use, we do not redistribute the underlying per-language EGIDS values. We release only the derived three-way classification, which is sufficient to reproduce this indicator.
        \item Written Status: from Ethnologue. To comply with Ethnologue reproducibility regulations, only report when a language is declared as written. 
    \end{itemize}

    \item \textbf{Digital infrastructure data}
    \begin{itemize}
        \item \% of households with: computers, phones, home internet — from ITU (\url{https://datahub.itu.int})
        \item \% of individuals using internet: (\url{https://data.worldbank.org/indicator/IT.NET.USER.ZS})
        \item Avg. Download/Upload Speeds, Latency (km$^2$ level): from Ookla \url{https://aws.amazon.com/marketplace/pp/prodview-breawk6ljkovm#overview}
    \end{itemize}
    
\end{itemize}
\end{enumerate}
All the collected existing data is summarised in Table \ref{tab:data_summary}.

\begin{table}[ht]
\centering
\caption{Summary of Extracted Variables, Their Sources and the literature behind them}
\begin{tabular}{p{4cm} p{6cm} p{5cm}}
\toprule
\textbf{Category} & \textbf{Variable} & \textbf{Source} \\
\midrule

\multirow{7}{*}{\textbf{AI Resources}} 
& Number of Speakers & Wikidata \url{https://www.wikidata.org/wiki/Help:Wikimedia_language_codes/lists/all} \\
& Number of Models & HuggingFace \url{https://huggingface.co/languages} \\
& Bible Translations & \url{https://www.bible.com/languages} \\
& CommonCrawl Data Volume & CommonCrawl \url{https://commoncrawl.github.io/cc-crawl-statistics/} \\
& OPUS Data Volume & OPUS \url{https://opus.nlpl.eu/} \\
& Wikipedia Data Volume & Wikipedia dumps \\
& Archival Resources & AILLA, ANLA, ELAR, DOBES, Pangloss, Kaipuleohone, PARADISEC, Wikitongues \\
& XEUS & \url{https://www.wavlab.org/activities/2024/xeus/}\\
& Keyboards & Keyman \url{https://keyman.com/en/}\\
& Level of Grammatical Documentation & Glottolog \url{https://glottolog.org/meta/glossary#sec-descriptivestatusofalanguage}\\

\midrule

\multirow{9}{*}{\textbf{Socioeconomic Data}} 
& Distance to Nearest University & 4ICU \url{https://www.4icu.org/reviews/index2.htm} \\
& Wikipedia Active Users & Wikidata \url{https://www.wikidata.org/wiki/Special:ActiveUsers}\\ %
& Language Ecology & Ethnologue and APiCS \url{https://www.ethnologue.com/} \& \url{https://apics-online.info/} \\
& Language Institutionality & Ethnologue \url{https://www.ethnologue.com/} \\
& Language Endangerment & UNESCO \url{https://en.wal.unesco.org/}\\
& Subnational Human Development Index (HDI) \citep{pick2021latin} & Global Data Lab \url{https://globaldatalab.org/shdi/} \\
& Subnational GDP per capita \citep{bagchi2005factors}& \citet{kummu2018gridded}, \url{https://github.com/mattikummu/griddedGDPpc} \\
& Subnational Educational Attainment \citep{hidalgo2020digital}& Global Data Lab \url{https://globaldatalab.org/shdi/} \\
& Cybersecurity Legislation Exists \citep{baliamoune2003analysis} & ITU \url{https://datahub.itu.int/data/?i=100103&s=8428} \\
& R\&D \% of GDP \citep{bagchi2005factors}& World Bank \url{https://data.worldbank.org/indicator/GB.XPD.RSDV.GD.ZS} \\
& Literacy Rate \citep{bagchi2005factors} & World Bank or UNESCO \\
& \% of STEM Graduates \citep{chetty2018bridging} & National Statistical Agencies / UNESCO \\

\midrule

\multirow{7}{*}{\textbf{Digital Infrastructure Data}} 
& \% of Households with Computers  \citep{baliamoune2003analysis} & ITU \url{https://datahub.itu.int/data/?i=12046} \\
& \% of Households with Phones \citep{itu2016measuring, harwit2004spreading, pick2015united, nishida2014japan} & ITU or World Bank \\
& \% of Households with Home Internet \citep{itu2016measuring} & ITU or World Bank \\
& \% of Individuals Using Internet \citep{loo2012developing, nishida2014japan, taylor2007measuring} & ITU or World Bank \\
& Avg. Download Speed (kbps) \citep{itu2016measuring, baliamoune2003analysis, nishida2014japan} & Ookla \url{https://aws.amazon.com/marketplace/pp/prodview-breawk6ljkovm} \\
& Avg. Upload Speed (kbps) \citep{itu2016measuring, baliamoune2003analysis, nishida2014japan}& Ookla \url{https://aws.amazon.com/marketplace/pp/prodview-breawk6ljkovm} \\
& Avg. Latency (ms) \citep{itu2016measuring, baliamoune2003analysis, nishida2014japan}& Ookla \url{https://aws.amazon.com/marketplace/pp/prodview-breawk6ljkovm} \\

\bottomrule
\end{tabular}
\label{tab:data_summary}
\end{table}
\clearpage
\subsection{Data exclusion}

No data exclusion 
\subsection{Missing data}

We addressed missing values through geographical interpolation: Let \( x_{r,c} \) represent the value of a variable for region \( r \) in country \( c \). Missing values are imputed according to the following rules:

\begin{itemize}
    \item \textbf{Subnational level interpolation:} 
    If \( x_{r,c} \) is missing for region \( r \), and at least one other region \( r' \) in the same country \( c \) has a non-missing value, then:
    \[
    x_{r,c} = \frac{1}{|R_c'|} \sum_{r' \in R_c'} x_{r',c}
    \]
    where \( R_c' \subset R_c \) is the set of regions in country \( c \) with available data.
    
    \item \textbf{National level interpolation:} 
    If all \( x_{r,c} \) are missing for country \( c \), then:
    \[
    x_{r,c} = x_{\text{similar}}
    \]
    where \( x_{\text{similar}} \) is the value from a country with a similar development level, determined by a composite development index (e.g., HDI or GDP per capita). We do not calculate this as a function of geographical proximity due to possibly very constrasting data (e.g. Comoros island, geographically closest to Mayotte (FR)). 
\end{itemize}

\section{Indicator justification}\label{lit}

We ground indicator selection in established digital-divide scholarship, ensuring that the index captures empirically recurrent determinants of technological disparity. Empirical research on global and subnational digital inequalities demonstrates that technological readiness emerges from the conjoint effects of socioeconomic development and institutional and infrastructural capacity. Classic cross-national analyses identify income, educational attainment, literacy, and governance quality as primary correlates of ICT diffusion \citep{bagchi2005factors, baliamoune2003analysis}, while work in digital geographies reveals persistent spatial clustering driven by structural asymmetries in human development, connectivity, and institutional investment \citep{pick2021latin, harwit2004spreading, pick2015united, nishida2014japan}. Indicators such as the Human Development Index \citep{pick2021latin}, GDP per capita \citep{bagchi2005factors, kummu2018gridded}, and educational attainment \citep{hidalgo2020digital} have repeatedly been shown to capture regional capacity and skill endowments underpinning digital participation \citep{chetty2018bridging}. Complementary governance indicators, including cybersecurity legislation, proxy the regulatory maturity and required for secure data flows and platform adoption \citep{baliamoune2003analysis, itu2016measuring}. R\&D expenditure, a marker of national innovation intensity, operates as a higher-order determinant of technological self-provisioning and the emergence of local computational capacity \citep{bagchi2005factors}. Spatial accessibility to universities further reflects regional knowledge infrastructures and technical labour markets, long recognized as accelerators of ICT diffusion \citep{pick2015united, nishida2014japan}.

Overall, infrastructure remains the most proximate constraint on digital engagement. Household-level access to computers, phones, and home internet—as monitored by ITU and the World Bank—continues to explain substantial variance in ICT use across developing and industrialized contexts \citep{itu2016measuring, loo2012developing}. High-throughput broadband deployment has increased reliance on quality-of-service metrics to evaluate the functional performance of digital infrastructure for advanced applications \citep{hidalgo2020digital}. Empirical network-performance data show that low-bandwidth, high-latency regions align with socioeconomic disparities, exacerbating structural digital divides \citep{pick2021latin}.

A parallel body of work on multilingual AI and linguistic resource allocation underscores the decisive role of digital content availability in determining technological readiness for specific languages. Modern language models depend on web-scale corpora (Common Crawl), parallel datasets (OPUS), and community-generated resources (Wikipedia); their extreme skew across languages remains a principal driver of performance disparities \citep{joshi2020state, blasi-etal-2022-systematic}. Historical translation traditions, particularly Bible translations, have provided high-quality parallel text for numerous low-resource languages and continue to underpin data scarcity mitigation strategies. The distribution of pretrained models in public repositories (e.g., Hugging Face \citep{huggingface_hub_docs_2023}) now functions as a direct measure of community attention and cumulative engineering investment, mirroring underlying socioeconomic and infrastructural inequalities.

Collectively, the evidence provided above converges on a multidimensional account of readiness in which socioeconomic endowments, institutional architectures, infrastructural quality jointly condition the feasibility of digital participation and the development of technologies.

\clearpage
\section{Supplementary figures}
\subsection{Validation of Huggingface data against ACL antology corpus}\label{validat
}

\begin{figure}[h!]
    \centering
    \includegraphics[width=\linewidth]{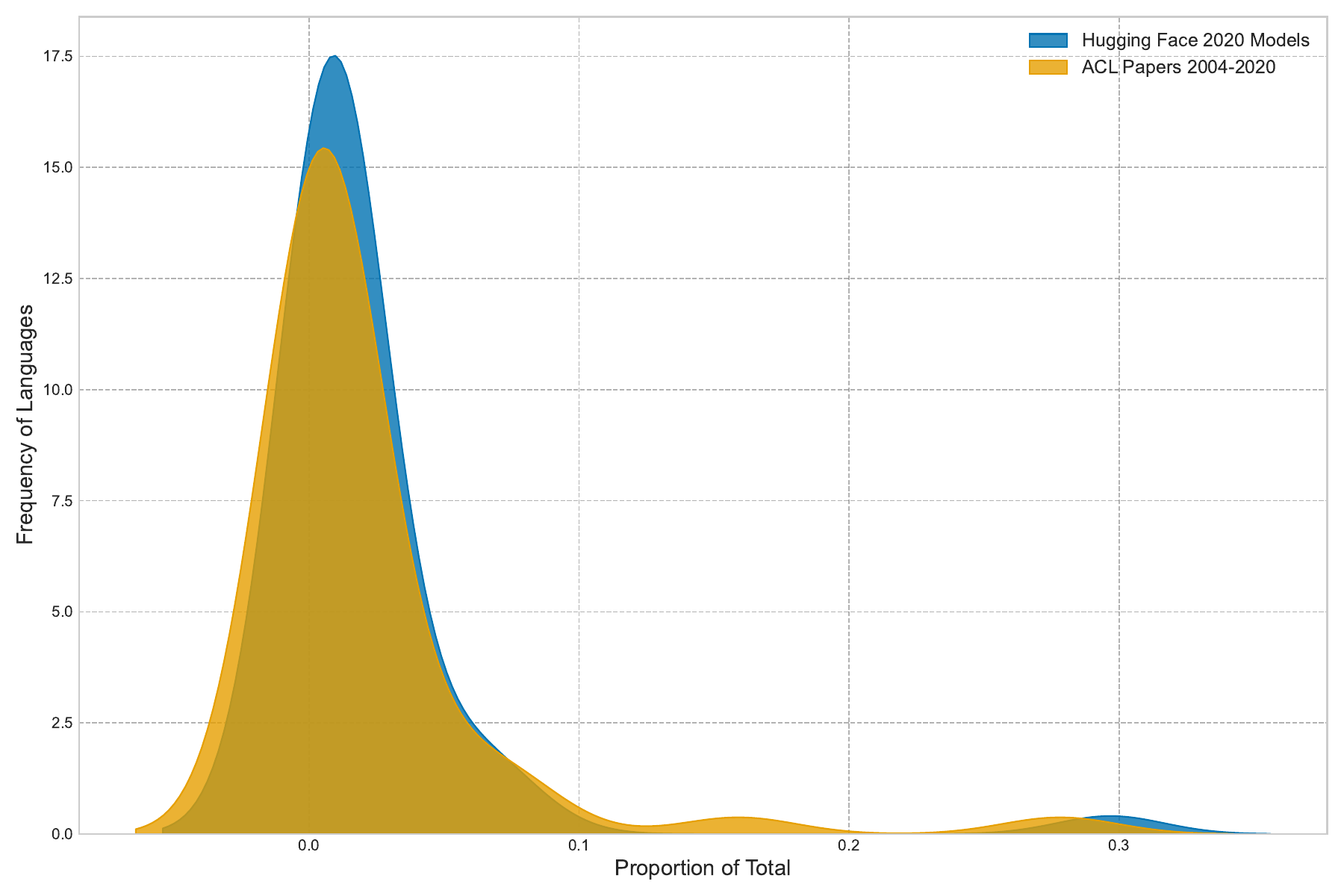}
    \caption{Kernel Density Estimate (KDE) plot showing the frequency density of language proportions for two distinct datasets. The Hugging Face 2020 Models (blue curve) represent the distribution of models available on the platform, while ACL Papers 2004-2020 (orange curve) shows the distribution of languages featured in the ACL anthology. The x-axis indicates the proportion of total language count, and the y-axis represents the frequency density.}
    \label{fig:valiud}
\end{figure}
\clearpage
\newpage
\subsection{Validation of findings against benchmarks }\label{benchs
}
\begin{figure}[h!]
    \centering
    \includegraphics[width=\linewidth]{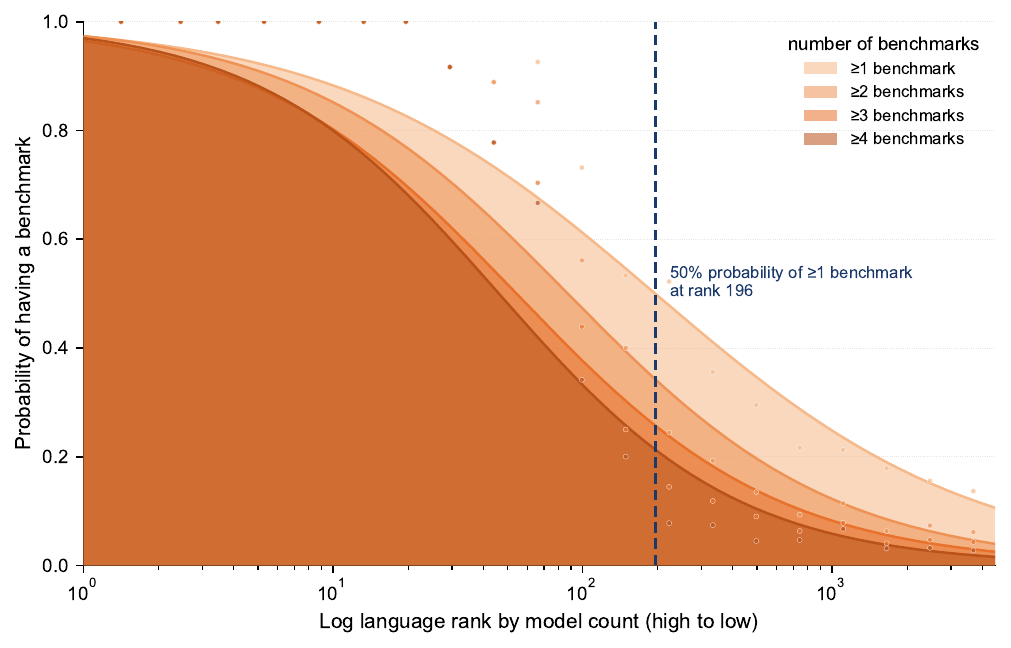}
\caption{Logistic fit of benchmark availability against language rank by model count. Each curve shows the modeled probability that a language has at least $n$ associated benchmarks ($n = 1$--$4$, light to dark orange), as a function of its rank in the model-count distribution ($x$-axis, log scale, highest to lowest). Curves are logistic regressions of benchmark presence on log-rank; points show the empirical proportion within logarithmically spaced rank bins. The probability of having any benchmark ($\geq 1$) falls below 50\% at rank 196 (dashed line), and the probability of having multiple benchmarks declines progressively faster, indicating that both benchmark availability and benchmark variety are concentrated among the most model-rich languages.}
    \label{fig:bench}
\end{figure}

\subsection{Validation of findings against monolinguality and multilinguality split}\label{monol
}
\begin{figure}[h!]
    \centering
    \includegraphics[width=\linewidth]{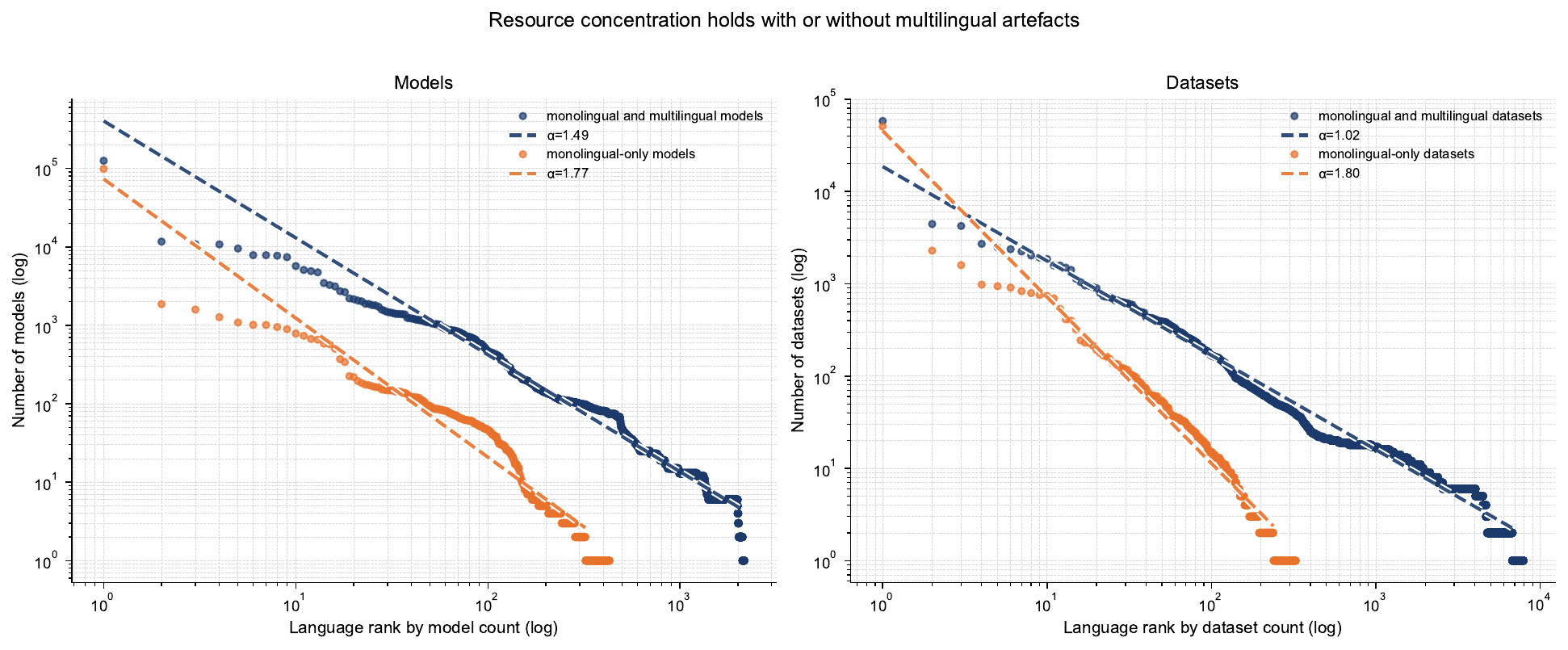}
\caption{Resource concentration among languages persists when multilingual artefacts are excluded. Rank--frequency distributions of Hugging Face artefacts per language, for models (left) and datasets (right), counting all artefacts (blue) and counting only those declared monolingual (orange). Both axes are logarithmic. Dashed lines are least-squares fits to the rank--frequency relation, with the fitted exponent $\alpha$.}
    \label{fig:monol}
\end{figure}
\subsection{Residuals of Zipf's models}
\begin{table}[h!]
    \centering
    \caption{Standard deviation of residuals ($\sigma_e$) from $\alpha=1$ Zipf's law fit per year for language models and datasets.}
    \label{tab:residuals_std_dev_combined}
    \begin{tabular}{|c|c|c|}
        \hline
        \textbf{Year} & \textbf{Models ($\sigma_e$)} & \textbf{Datasets ($\sigma_e$)} \\
        \hline
        2020 & 0.417 & 0.365 \\
        2021 & 0.549 & 0.716 \\
        2022 & 0.878 & 0.743 \\
        2023 & 1.130 & 0.801 \\
        2024 & 0.668 & 0.133 \\
        \hline
    \end{tabular}
\end{table}
\newpage

\subsection{Geographical distribution of resources}\label{geo_bigger}
\begin{table}[h!]
    \centering
    \caption{Sensitivity of the speaker--model correlation to exclusion of the most-spoken languages. Log--log OLS refit of the number of models against speaker population after removing the top $k$ languages by speaker population. The correlation declines only modestly and remains moderate and highly significant throughout, indicating that the relationship is not an artefact of the most-resourced languages.}
    \label{tab:sensitivity}
    \begin{tabular}{cccccc}
        \toprule
        \textbf{Top $k$ removed} & \textbf{$n$} & \textbf{Slope} & \textbf{$R^2$} & \textbf{Pearson $r$} & \textbf{$p$} \\
        \midrule
        0  & 1732 & 0.312 & 0.304 & 0.551 & $2.6\times10^{-138}$ \\
        5  & 1727 & 0.299 & 0.286 & 0.535 & $3.3\times10^{-128}$ \\
        6  & 1726 & 0.297 & 0.283 & 0.532 & $1.1\times10^{-126}$ \\
        7  & 1725 & 0.296 & 0.280 & 0.530 & $2.6\times10^{-125}$ \\
        10 & 1722 & 0.289 & 0.272 & 0.522 & $9.9\times10^{-121}$ \\
        15 & 1717 & 0.288 & 0.268 & 0.517 & $4.4\times10^{-118}$ \\
        20 & 1712 & 0.285 & 0.263 & 0.513 & $1.7\times10^{-115}$ \\
        \bottomrule
    \end{tabular}
\end{table}
\begin{figure}[h!]
    \centering
    \includegraphics[width=\linewidth]{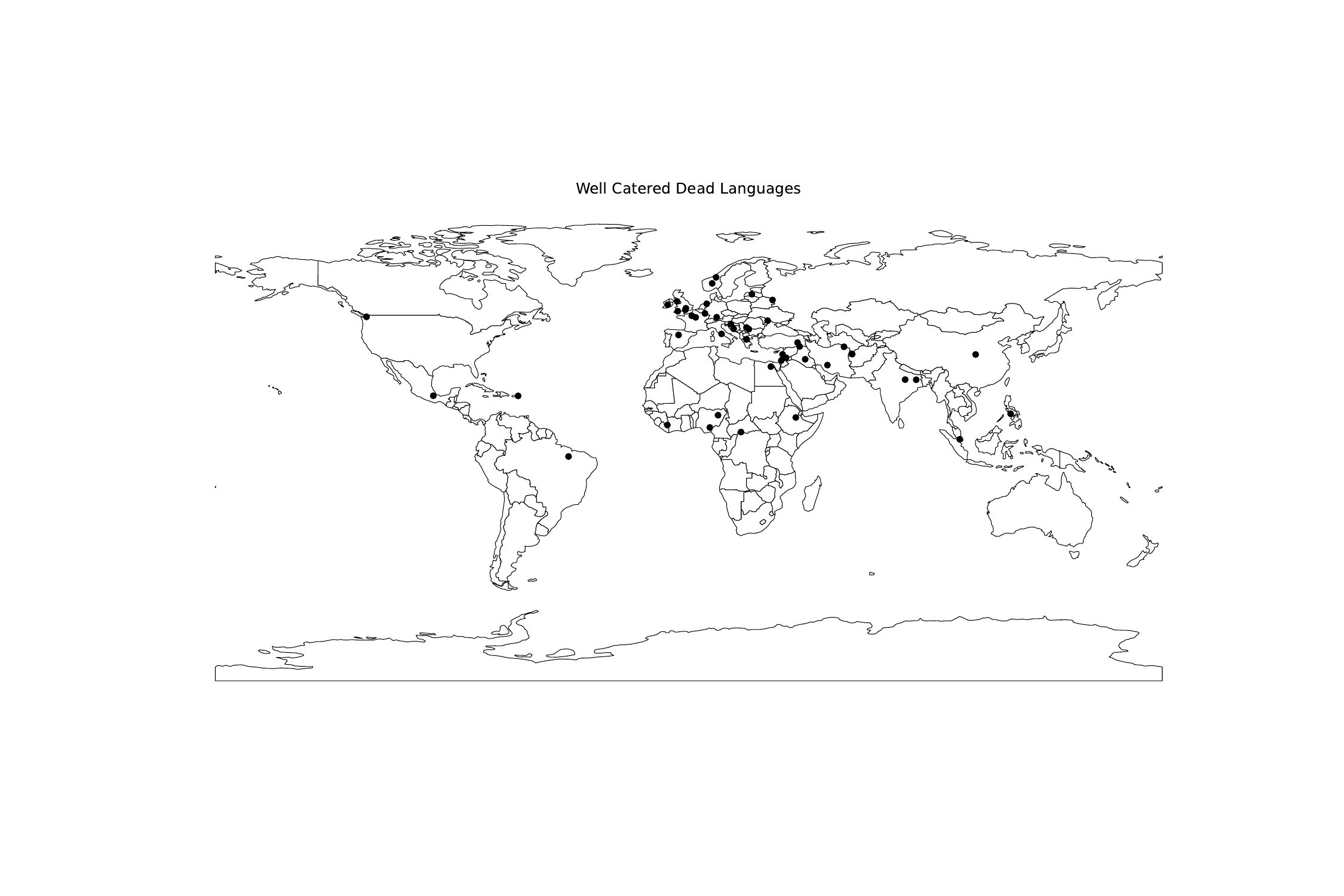}
    \caption{The vast majority of dead languages with models and datasets are located in Europe. }
    \label{fig:deads}
\end{figure}

\begin{figure}[h!]
    \centering
    \includegraphics[width=\linewidth]{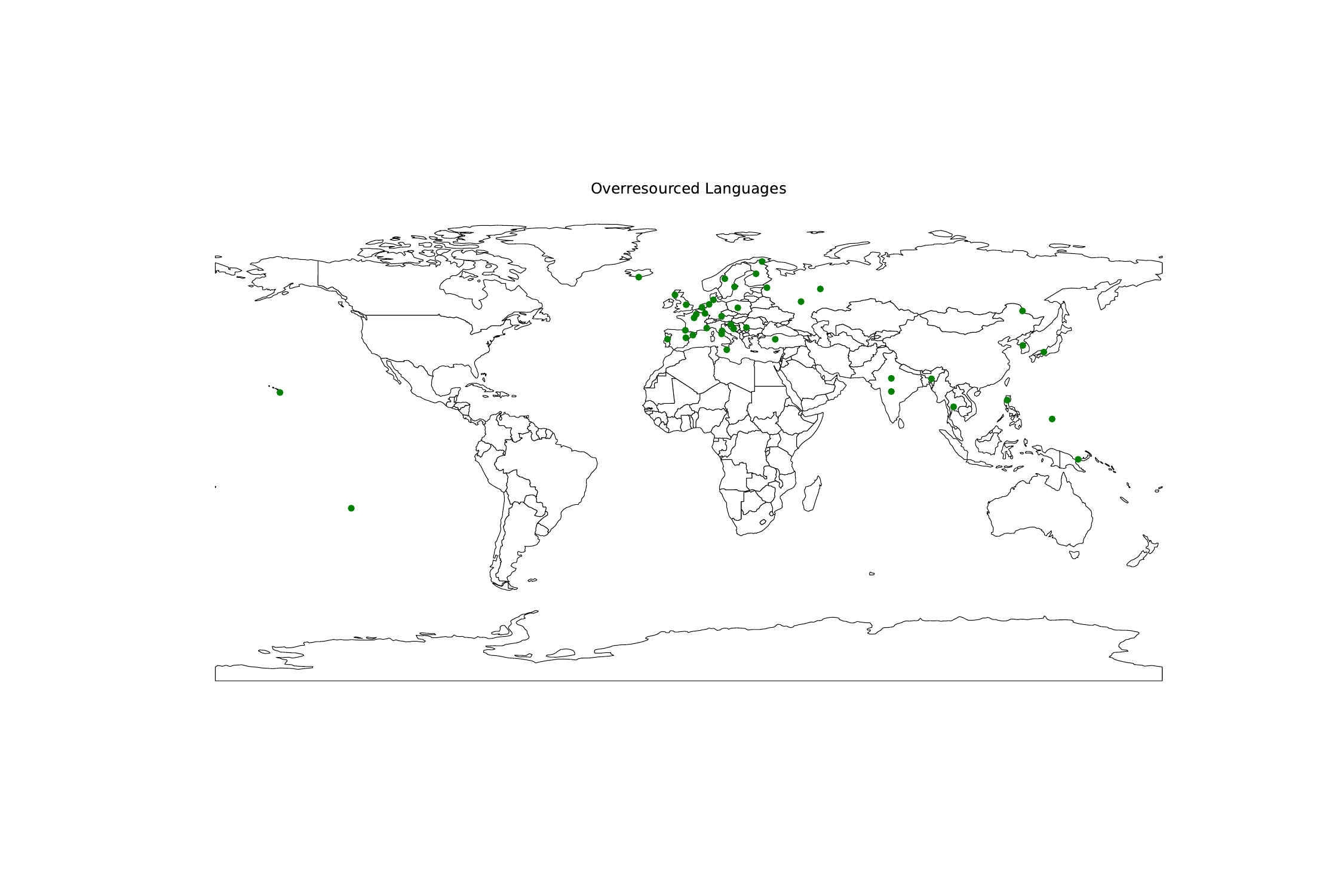}
    \caption{The majority of overresourced languages with models and datasets are located in Europe.}
    \label{fig:overr}
\end{figure}

\begin{figure}[h!]
    \centering
    \includegraphics[width=\linewidth]{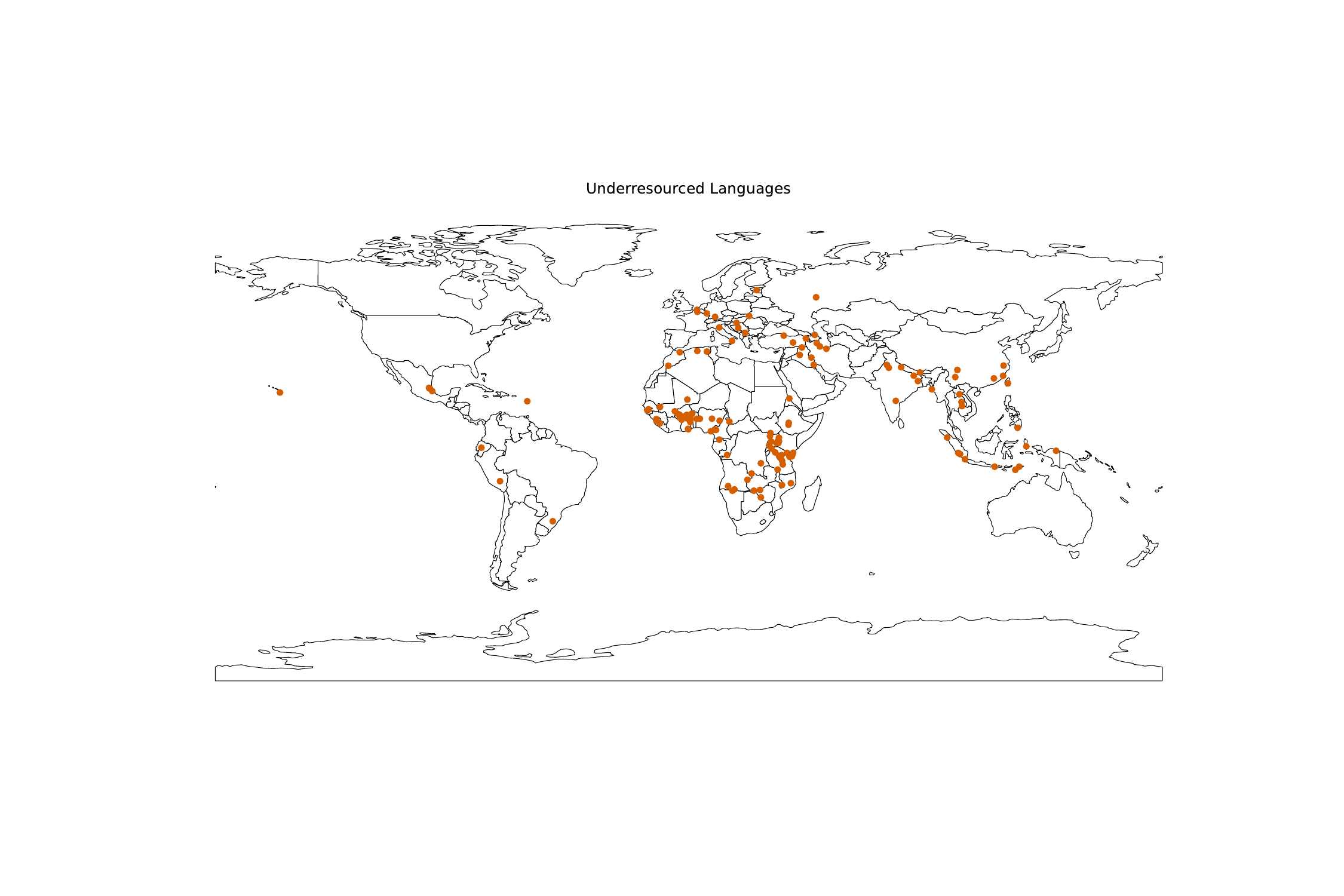}
    \caption{The vast majority of underresourced languages with models and datasets are located in Africa.}
    \label{fig:underr}
\end{figure}

\begin{figure}[h!]
    \centering
    \includegraphics[width=\linewidth]{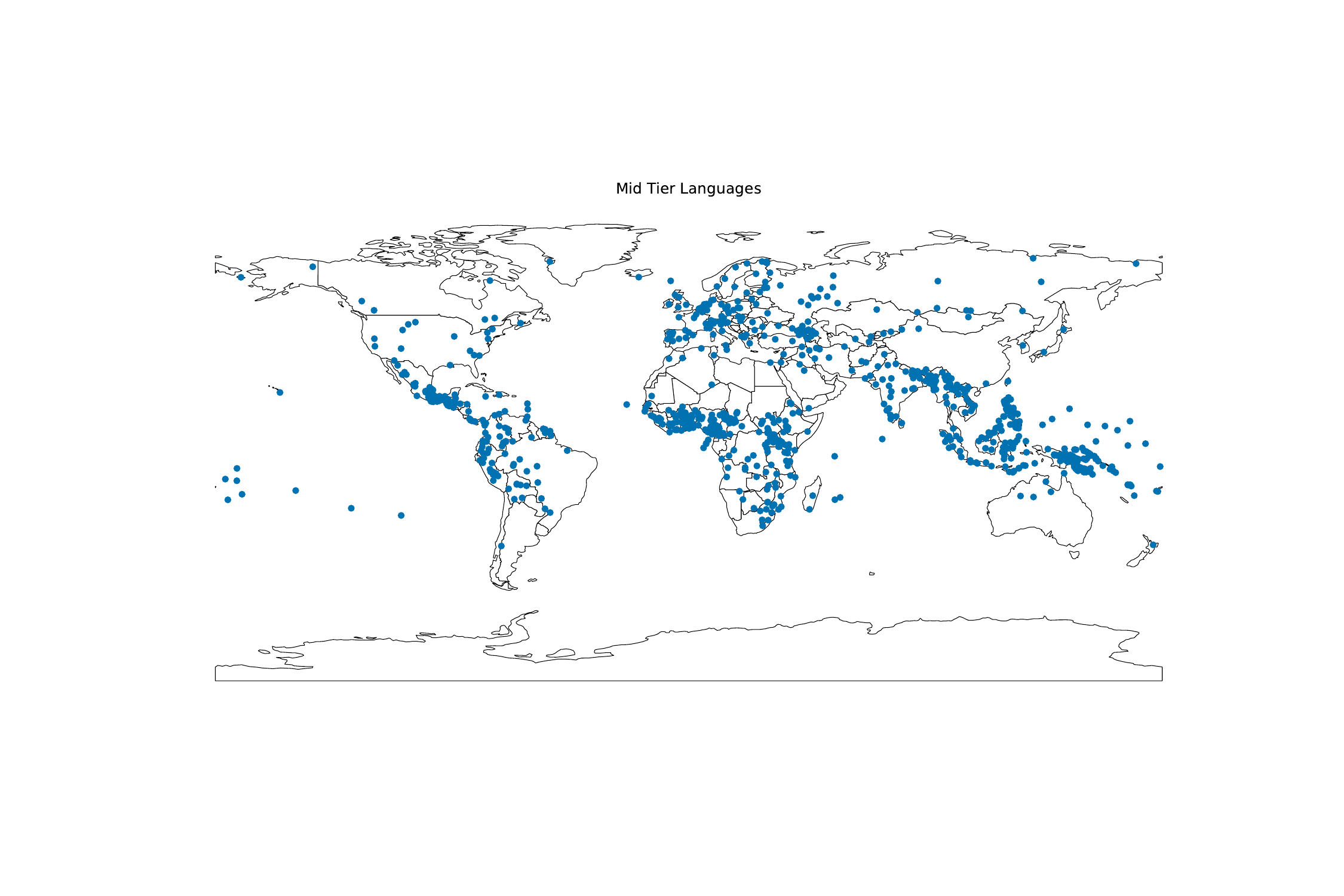}
    \caption{Distribution of mid-tier languages, the intermediate group with model presence, comprising those not classified as dead, under-resourced, or among the top five per speaker bin. They span a wide range of speaker populations and are distributed across all inhabited continents.}
    \label{fig:mid_tier}
\end{figure}

\begin{figure}[h!]
    \centering %

    \begin{minipage}{0.48\textwidth} %
        \centering
        \includegraphics[width=\linewidth]{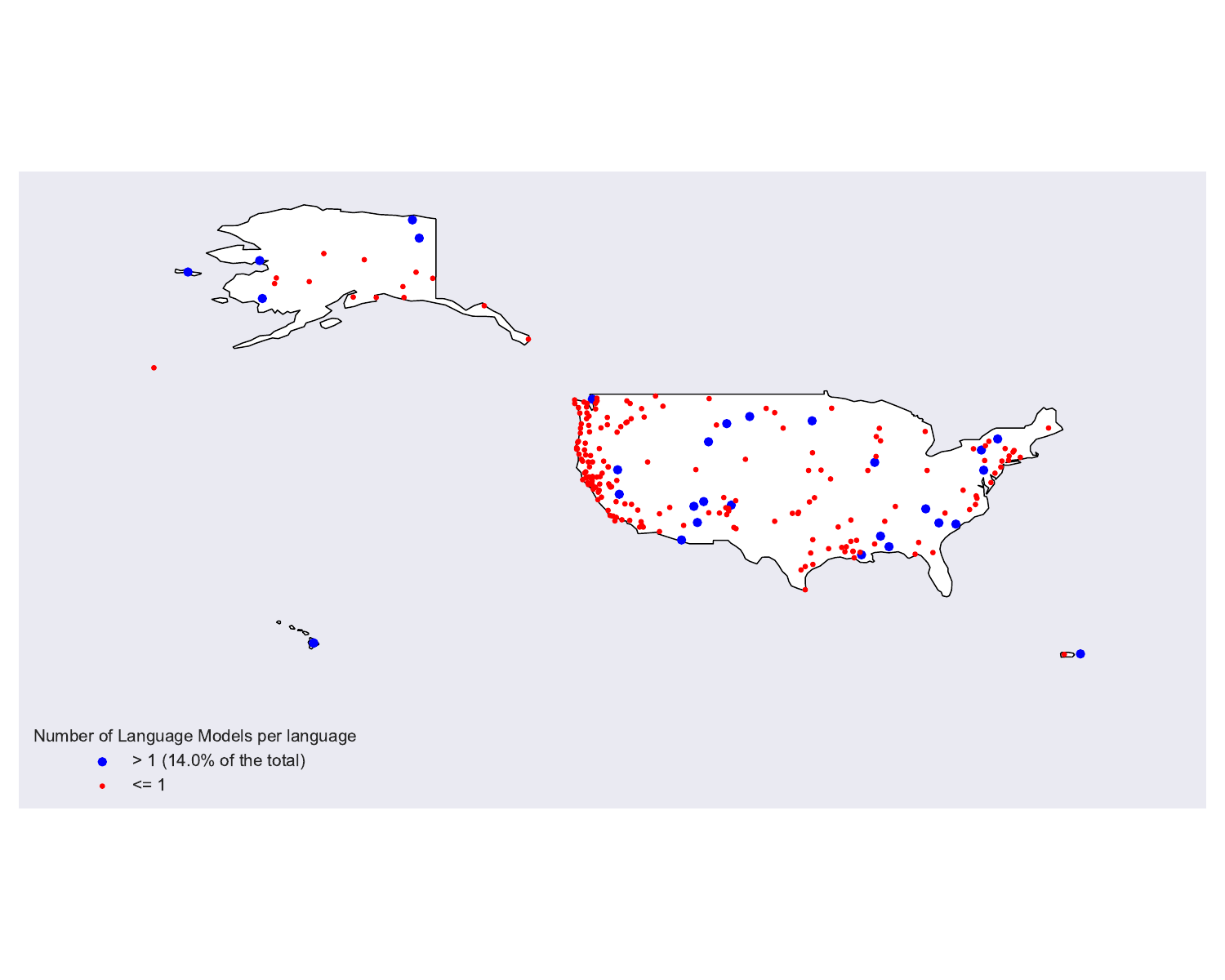}
    \end{minipage}
    \hfill %
    \begin{minipage}{0.48\textwidth} %
        \centering
        \includegraphics[width=\linewidth]{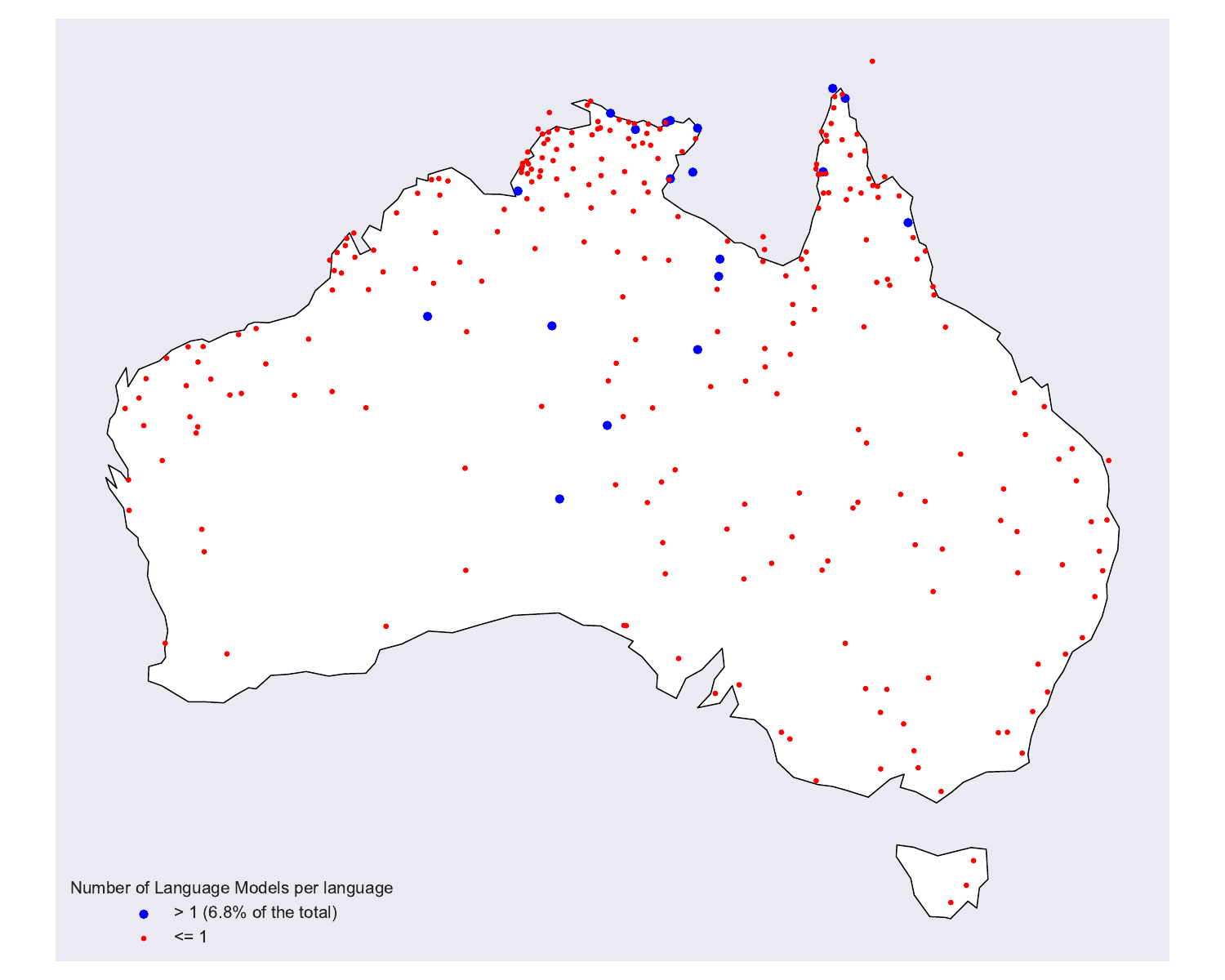}
    \end{minipage}

    \vspace{1em} %

    \begin{minipage}{0.48\textwidth} %
        \centering
        \includegraphics[width=\linewidth]{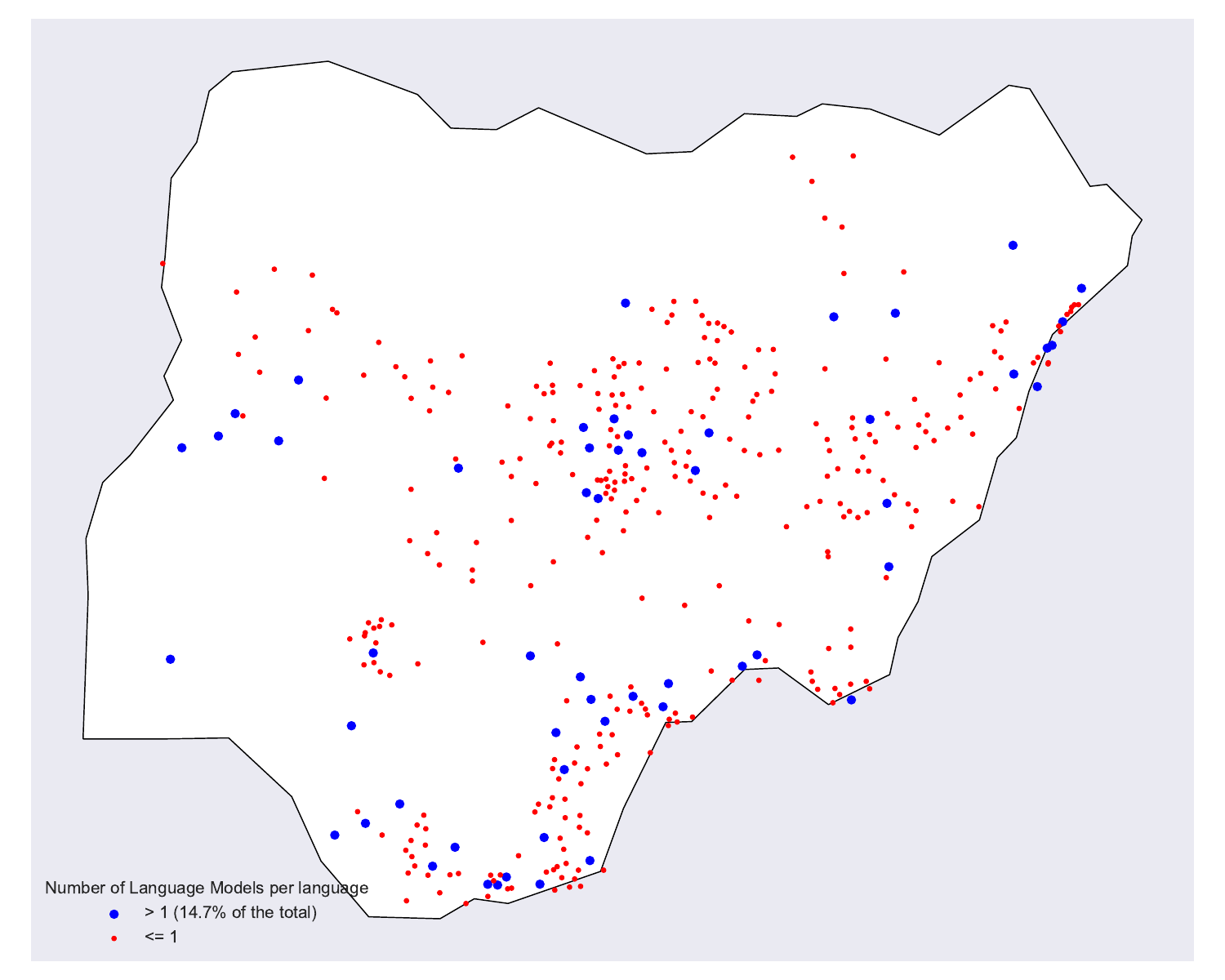} %
    \end{minipage}

    \caption{Comparative visualization of language technology development across the United States, Australia, and Nigeria reveals an unexpected pattern. Nigeria, a comparatively less technologically developed country, has a higher proportion of languages covered by at least one language model (14.7\%) than both the United States (14\%) and Australia (6.8\%). Paradoxically, the region with the largest number of languages left behind in the United States is California—the most technologically advanced area in the world.}
    \label{fig:combined_images_3}
\end{figure}

\clearpage
\newpage

\subsection{Technological diffusion parameters}\label{gompertz_params}

\begin{figure}[h!]
    \centering
    \includegraphics[width=\linewidth]{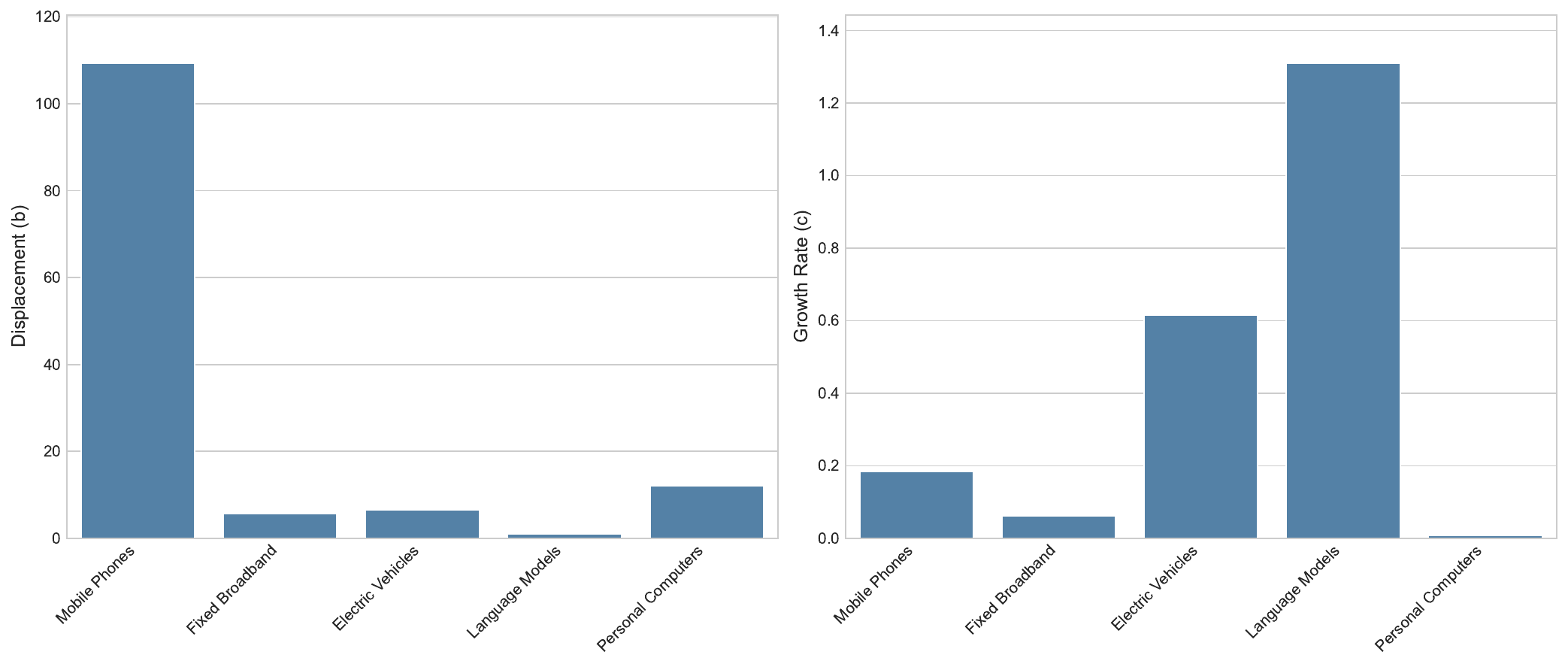}
    \caption{The Gompertz model fitted to the adoption rate of various technologies. The left panel shows the displacement parameter ($b$), which indicates the horizontal shift of the growth curve. The right panel displays the growth rate constant ($c$), which represents the steepness and speed of adoption. Language Models exhibits the highest growth rate and the lowest displacement, indicating a very rapid, early-stage adoption trajectory.}
    \label{fig:combined_gompz_plots_with_legend}
\end{figure}

\subsection{Data analysis}

\begin{figure}[h!]
    \centering
    \includegraphics[width=\linewidth]{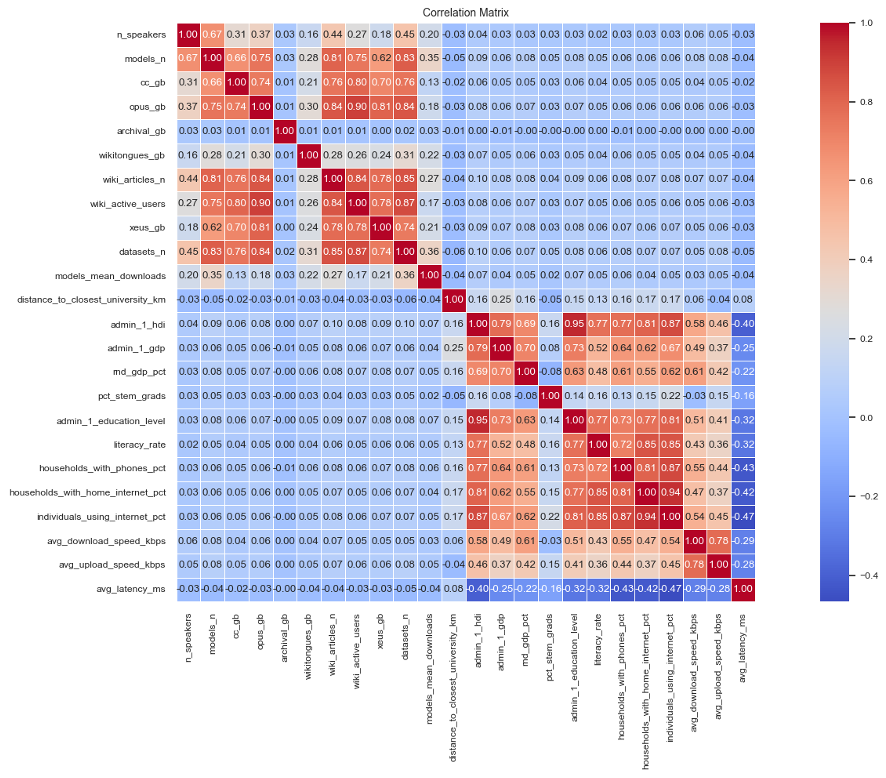}
    \caption{Heatmap illustrating the pairwise Pearson correlation coefficients between 24 variables, ordered to display the distinct clustering of socioeconomic and  digital infrastructure indicators on the one hand, and AI resources indicators on the other. The color scale, ranging from deep blue (strong negative correlation) to deep red (strong positive correlation), reveals two highly correlated internal blocks: one comprising AI resource availability indicators (e.g., cc\_gb, datasets\_n) and the other comprising socioeconomic and digital infrastructure indicators (e.g., admin\_1\_hdi, literacy\_rate, avg\_download\_speed\_kbps). Crucially, the coefficients across these two blocks are predominantly weak (near zero, white/light grey), quantitatively showcasing the structural independence of general societal development from the specific availability of AI-relevant resources.}
\label{fig:corr_matrix}
\end{figure}

\begin{figure}[h!]
    \centering
    \includegraphics[width=\linewidth]{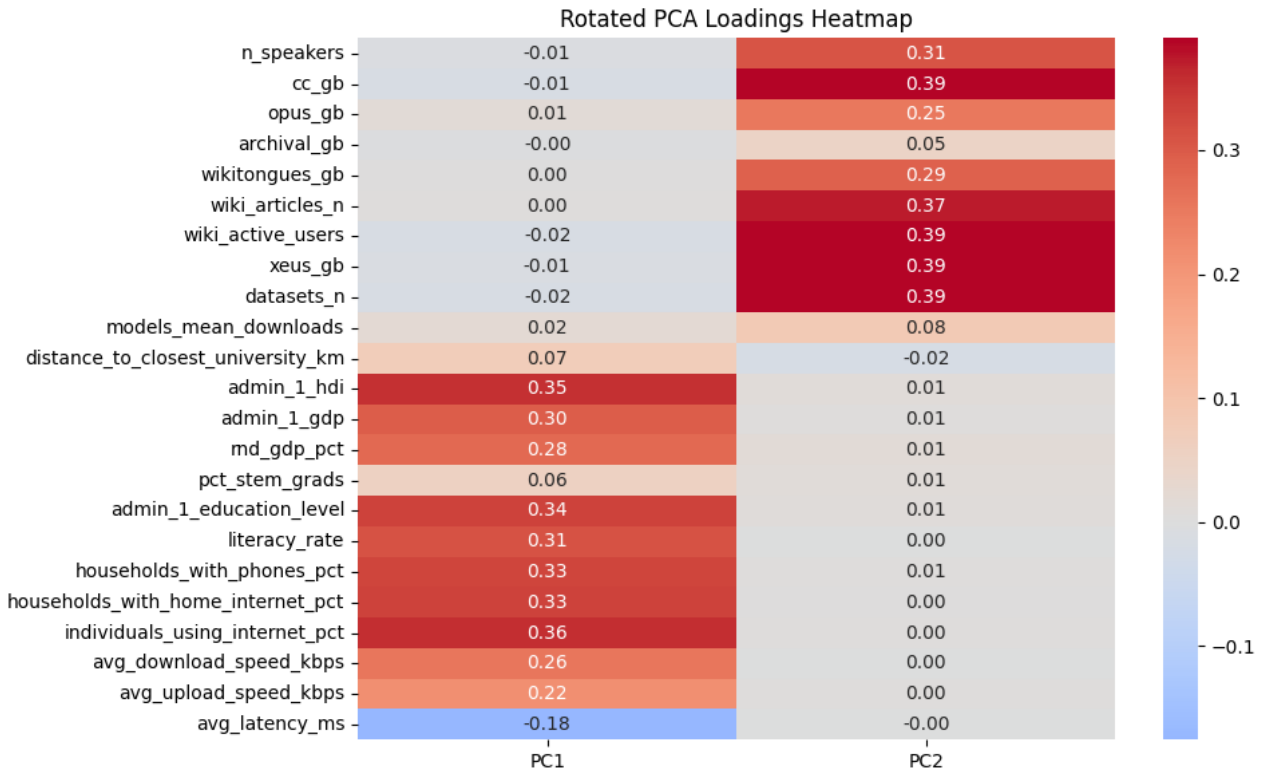}
    \caption{Heatmap illustrating the rotated component loadings for 24 variables on the first two principal components (PC1 and PC2) following Varimax rotation. PC1 is defined by high positive loadings from social and infrastructural indicators, including the proportion of individuals using the internet (loading =0.35), Human Development Index (HDI; loading =0.35), and education level (loading =0.33). Conversely, PC2 is strongly characterized by high positive loadings for AI technology availability, with the highest values observed for the volume of data on CommonCrawl (loading =0.37), the number of datasets on HuggingFace (loading =0.37), and Wikipedia active users (loading =0.37). The color intensity reflects the magnitude of the loading, ranging from a strong negative correlation (deep blue, e.g., avg\_latency\_ms on PC1) to a strong positive correlation (deep red). Together, these two orthogonal components explain 58.4\% of the total variance, revealing a clean separation between foundational societal development and the specific availability of AI-focused resources.}
\label{fig:pca_1}
\end{figure}

\begin{figure}[h!]
    \centering
    \includegraphics[width=\linewidth]{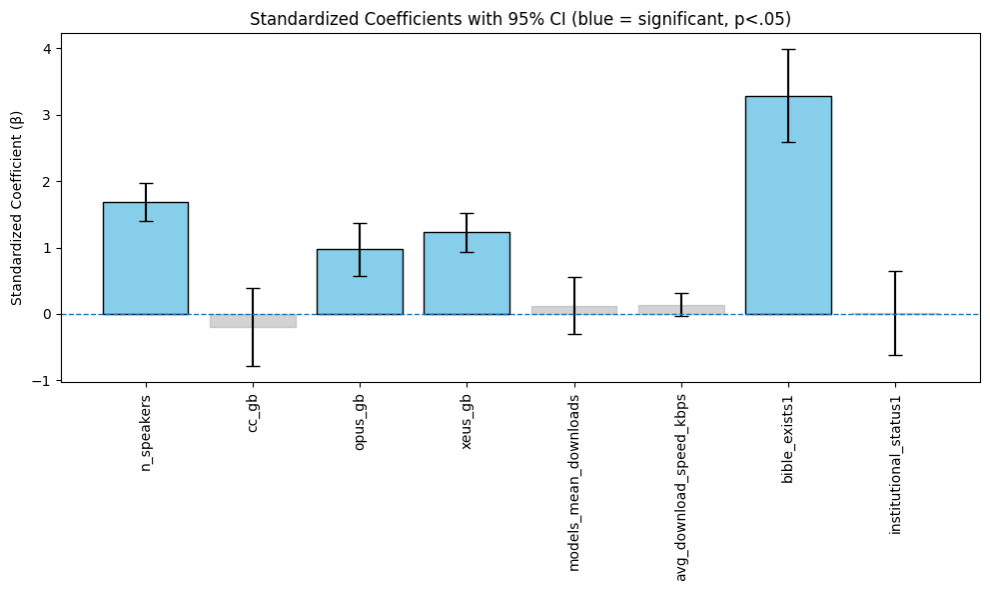}
    \caption{Standardized Regression Coefficients ($\beta$) from Stepwise Model Selection. Bar plot displaying the standardized regression coefficients ($\beta$) and $95\%$ confidence intervals (CI) for predictors retained in a final model after stepwise regression. Coefficients colored light blue are statistically significant ($p < 0.05$), indicating a reliable positive relationship with the dependent variable. Significant positive predictors include the number of speakers ($n\_speakers$), the volume of Opus data ($opus\_gb$), the volume of Xeus data ($xeus\_gb$), and the strongest predictor by magnitude, the presence of a Bible translation ($bible\_exists1$). Variables in light gray (e.g., $cc\_gb$, $models\_mean\_downloads$) were retained but found to be statistically non-significant ($p \ge 0.05$), as indicated by the $95\%$ CI overlapping the zero line (dashed blue). The standardized nature of the coefficients allows for a direct comparison of the relative predictive strength of each significant variable.}
\label{fig:stepwise}
\end{figure}

\newpage
\subsection{Surveys output}
\subsubsection{Respondents characteristics}\label{survey_participants}

\begin{table}[!htbp]
    \centering
    \caption{Affiliations and roles of participants in Survey 1, the ranking of AI resources. Respondents were experts in multilingual language technology (n = 28).}
    \label{tab:institution_counts_NLP}
    \begin{tabular}{lllc}
        \toprule
        \textbf{Institution/Organization} & 
        \textbf{Country} & \textbf{Role} &\textbf{Count}\\
        \midrule
        LMU Munich & Germany & PhD Student & 5 \\
 
        Uppsala University & Sweden & Professor & 1 \\
        University of Cambridge & United Kingdom & Postdoc & 1 \\
        Würzburg University & Germany & Professor & 1 \\
        McGill University & Canada & Professor & 1\\
        Google DeepMind & United Kingdom  & Research Scientist &1\\
        Cohere For AI & Canada \& USA & Research Scientist & 2\\
        Carnegie Mellon University & USA & Professor & 1\\
        Meta & USA & Research Scientist & 1 \\
        Stanford University & USA & PhD Student & 1 \\
        MBZUAI & UAE & Professor & 1\\
        University of Washington & USA & PhD Student & 1 \\
        George Mason University & USA & Professor & 1 \\
        Swansea University & United Kingdom & Professor & 1 \\
        University of Colorado Boulder & USA & PhD Student & 1 \\
        University of the Basque Country & Spain & Professor & 1 \\
        Microsoft Research Labs India & India & Researcher & 2 \\
        École Normale Supérieure & France & Professor & 1\\
        University of Latvia & Latvia & Professor & 1 \\
        UNESCO Nigeria & Nigeria & Consultant & 1 \\
        UNESCO Wales & United Kingdom & Policy Maker & 1 \\
        Technion  & Israel & Professor & 1 \\
        \midrule
        \multicolumn{2}{l}{\textbf{Total}} & & \textbf{28} \\
        \bottomrule
    \end{tabular}
\end{table}
 
\begin{table}[h!]
    \centering
    \caption{Affiliations and roles of participants in Survey 2, the ranking of digital-infrastructure and socioeconomic factors. Respondents were experts in human–computer interaction and global technology adoption (n = 20).}
    \label{tab:institution_counts_non_NLP}
    \begin{tabular}{lllc}
        \toprule
        \textbf{Institution/Organization} & 
        \textbf{Country}& \textbf{Role} & \textbf{Count} \\
        \midrule
        University of Cambridge & United Kingdom & Postdoctoral Researcher &2 \\
        University of Bristol & United Kingdom & Professor & 2\\
        Cohere For AI & Canada \& USA & Research Scientist & 2\\
        Stanford University & USA & PhD Student & 1 \\
        University of Washington & USA & PhD Student &  1\\
        George Mason University & USA & Professor & 1\\
        Swansea University & United Kingdom & Professor & 1  \\
        University of Colorado Boulder & USA & PhD Student & 1 \\
        Imperial College London & United Kingdom & Professor and PhD Student &3 \\
        UiT The Arctic University of Norway & Norway & Professor & 1 \\
        Microsoft Research Labs India & India & Researcher & 2 \\
        University of Latvia & Latvia & Professor & 1 \\
        UNESCO Nigeria & Nigeria & Consultant & 1 \\
        UNESCO Wales & United Kingdom & Policy Maker & 1  \\
        \midrule
        \multicolumn{2}{l}{\textbf{Total}} & & \textbf{20} \\
        \bottomrule
    \end{tabular}
\end{table}
\newpage
\subsubsection{Survey results}\label{surveyr_results}
\begin{table}[h]
    \caption{Ranked AI resources in order of importance for building a language technology system.}
    \begin{tabular}{ll}
        \toprule
        \textbf{Rank} & \textbf{Resource} \\
        \midrule
        1 & Amount of data in Common Crawl (bytes)\\
        2 & Amount of data in Wikipedia Articles (bytes) \\
        3 & Number of already available multilingual Language Models  \\
        4 & Typological similarity to well resourced language families\\
        5 & Amount of data in OPUS (bytes) \\
        6 &  Number of already available monolingual Language Models \\
        7 & Existence of a dictionary\\
        8 & Existence of a Bible \\
        9 & Other Archival Data (bytes) e.g. linguistic documentation archives\\
        10 & Amount of audio data (in bytes) \\
        \bottomrule
    \end{tabular}
\end{table}
\begin{table}[htbp]
    \centering
    \caption{Ranked digital infrastructure factors in order of importance for bridging digital divides that affect the possibility of development and long-term sustainability of language technologies in under-resourced communities.}
    \begin{tabular}{ll}
        \toprule
        \textbf{Rank} & \textbf{Resource} \\
        \midrule
        1 & Percentage of households with phones \\
        2 & Percentage of individuals with home internet \\
        3 & Average network speed (upload, download, latency)  \\
        4 & Percentage of households with home internet access \\
        5 & Existence of a nearby university or technical institution \\
        6 &  Availability of local cloud computing infrastructure (e.g., data centers, regional hosting services)\\
        7 & Existence of Cybersecurity legislation/regulation\\

        \bottomrule
    \end{tabular}
\end{table}
\begin{table}[htbp]
    \centering
    \caption{Ranked socioeconomic factors in order of importance for bridging digital divides that affect the possibility of development and long-term sustainability of language technologies in under-resourced communities.}
    \begin{tabular}{ll}
        \toprule
        \textbf{Rank} & \textbf{Resource} \\
        \midrule
        1 & 	Language community's Institutional Status \\
        2 & Number of Speakers \\
        3 & 	Language community's Vitality Status \\
        4 & 	Online Active Wikipedia Users (Digital Vitality) \\
        5 & 	Literacy Rate \\
        6 &  Gross Domestic Product \\
        7 & Research and Development Spending (\% of GDP)\\
        8& Human Development Index\\
        9& Educational Attainment Score \\
        10& Closeness to Universities and Research Centers\\
        \bottomrule
    \end{tabular}
\end{table}

\clearpage

\end{appendices}

\clearpage

\end{document}